%% file: MAIN.tex
\documentclass[aps,physrev,preprint,superscriptaddress]{revtex4-2}
\usepackage{graphicx}
\usepackage{epstopdf, epsfig}
\usepackage{mdframed}
\usepackage{nomencl}
\usepackage{booktabs}
\usepackage{afterpage}
\usepackage[utf8]{inputenc}
\usepackage{siunitx}
\usepackage{comment}
\usepackage{booktabs}  
\usepackage{float}
\usepackage{soul}
\usepackage{caption}
\usepackage{acronym}
\usepackage{tablefootnote}
\usepackage[flushleft]{threeparttable}
\usepackage{mathtools,tikz,caption}
\DeclareRobustCommand\sampleline[1]{%
  \tikz\draw[#1] (0,0) (0,\the\dimexpr\fontdimen22\textfont2\relax)
  -- (2em,\the\dimexpr\fontdimen22\textfont2\relax);%
}
\acrodef{DNS}[DNS]{Direct Numerical Simulations}
\acrodef{IR}[IR]{InfraRed}
\acrodef{NETD}[NETS]{Noise Equivalent Temperature Difference}
\acrodef{PIV}[PIV]{Particle Image Velocimetry}
\acrodef{POD}[POD]{Proper Orthogonal Decomposition}
\acrodef{TBL}[TBL]{turbulent boundary layer}
\acrodef{ZPG}[ZPG]{zero-pressure gradient}

\begin{document}

\preprint{ }

\title{Instantaneous convective heat transfer at the wall: a depiction of turbulent boundary layer structures}

\author{Firoozeh Foroozan}
\email{Contact author: fforooza@ing.uc3m.es}
\altaffiliation{Department of Aerospace Engineering, Universidad Carlos III de Madrid, 28911 Leganés, Spain}

\author{  Andrea Ianiro}
\affiliation{Department of Aerospace Engineering, Universidad Carlos III de Madrid, 28911 Leganés, Spain}

\author{  Stefano  Discetti}
\affiliation{Department of Aerospace Engineering, Universidad Carlos III de Madrid, 28911 Leganés, Spain}

\author{  Woutijn J. Baars}
\affiliation{Faculty of Aerospace Engineering, Delft University of Technology, 2629 HS Delft, The Netherlands}

\date{\today}
\begin{abstract}
\vspace{0.2cm}

\noindent We demonstrate the ability to experimentally measure fluctuations of the convective heat transfer coefficient at the wall in a turbulent boundary layer. For this, we measure two-dimensional fields of wall-temperature fluctuations beneath a zero-pressure-gradient turbulent boundary layer, at two moderate friction Reynolds numbers ($Re_\tau \approx 990$ and $Re_\tau \approx 1800$). Spatiotemporal data of wall-temperature are acquired by means of a heated-thin-foil sensor as sensing hardware, and an infrared camera as temperature detector. At low $Re_\tau$ conditions, the fields of the Nusselt number fluctuations are populated by elongated structures comprising streamwise and spanwise length scales comparable to those of near-wall streaks. At higher $Re_\tau$ conditions, the effective width and length of the coherent $Nu$ fluctuations increases. These findings are based on two-point correlations, as well as streamwise-spanwise energy spectra of $Nu$ fluctuations. The convective velocities of the $Nu$ fluctuations are also computed with the available time resolution from the measurements. This allows for resolving the multi-scale nature of convective footprints of wall-bounded turbulence: our experimental data reflect that larger streaks in the footprint convect at velocities in the order of the free-stream velocity, while the more energetic smaller-scale features move at velocities in the order of $10u_\tau$. Measurements of the kind presented here offer a promising method for sensing, as they can be used as input to flow control systems. 
\begin{description}
\item[keywords]
Turbulent boundary layers, Non-intrusive sensing, Infrared thermography, Convective heat transfer, Turbulent convection, Boundary layer structure.
\end{description}
\end{abstract}

\maketitle
    \input{Sections/1-Introduction}
    \input{Sections/2-Experimental_Setup}
    \input{Sections/3-Wall_Measurement}
    \input{Sections/3.1}
    \input{Sections/3.2}
    \input{Sections/3.3}
    \input{Sections/4-Conclusions}

\section*{ACKNOWLEDGEMENT}

\noindent This activity was supported by the project ARTURO, funded by the Spanish State Research Agency, ref.PID2019-109717RB-I00/AEI/10.13039/501100011033, and by the project EXCALIBUR (Grant No PID2022-138314NB-I00), funded by MCIU/AEI/10.13039/501100011033 and by “ERDF A way of making Europe”. \\
F. Foroozan was partially supported by the UC3M mobility program.\\
We thank R. Castellanos for assisting with the design of the heated-thin-foil sensor. We also express gratitude to the technicians of the Flow Physics \& Technology laboratory within the Faculty of Aerospace Engineering, at the Delft University of Technology, for assistance in setting up the experiments. Finally, we acknowledge G. Dacome and F. Schrijer for contributing to the characterization of the TBL flow and to the acquisition setup of the IR camera, respectively.

\section*{}

\bibliography{references.bib}
\end{document}

%% file: Sections/1-Introduction.tex
\section{Introduction}
\label{sec:Intro}

Viscous interactions between turbulence and solid boundaries directly affect mass, momentum and energy transport. Understanding, modeling and accurately sensing the dynamics of \acp{TBL} are essential for the performance of a wide range of engineering devices. For instance, active control of \acp{TBL} requires fast, accurate and non-intrusive sensing strategies, thus leaving the paradigm of sensing-from-the-wall as the most appealing option \citep{cattafesta2011actuators}. Our current work focuses on the feasibility of performing time-resolved convective heat transfer measurements in a TBL flow with air as the working fluid, for the purpose of inferring the spatiotemporal footprint of the convecting flow. We rely on time-resolved \ac{IR} thermography, coupled with a heated-thin-foil heat transfer sensor \citep{astarita2012infrared}.

Being able to measure the footprint of wall-bounded turbulence in a non-intrusive manner is an essential step towards implementing real-time control of wall-bounded flows. 
Real-time control of off-the-wall velocity structures relies on methodologies to estimate instantaneous turbulent velocity fluctuations, often based on the availability of wall data alone \citep[\emph{e.g.},][]{guemes2019sensing,encinar2019logarithmic, guemes2021coarse,guastoni2021convolutional,cuellar2024three}. 
Most methods require instantaneous fields of both the stream-wise and span-wise components of skin friction, and pressure. While those data are easily accessible in numerical simulations, experimentally measuring these three quantities over a grid of points is challenging. Fortunately, recent works suggest that only one quantity is sufficient for meaningful estimates of the off-the-wall velocity fields \citep{cuellar2024some,guastoni2024fully}, as also demonstrated in experimental studies on real-time control of TBL flow \citep{Dacome2024} and on turbulence estimation from intrusive and sparse measurements \citep[][among others]{kerherve2017combining, discetti2019characterization}.


Typically, the wall-footprint of turbulence dynamics can be obtained through measurements of the unsteady wall-pressure or wall-shear stress. Low values of turbulence-induced pressure fluctuations yield a challenge for the measurement of instantaneous wall-pressure fields \citep{pastuhoff2013enhancing,gu2024denoising}. Likewise, challenges exist in the measurement of wall-shear-stress fields, \emph{e.g.,} with a deformable sensor whose inertia limits the maximum measurement frequency \citep{amili2011film}. A convective heat flux sensor offers an alternative to measuring wall-shear stress, and is considered in the current work. The similarity between mean momentum and energy transport in wall-bounded flows was first discussed by \citet{reynolds1874extent} \citep[later reproduced in][]{reynolds1961extent}. Beyond the range of validity of the Reynolds analogy, it is rather well-known that a strong correlation exists between turbulent heat transfer fluctuations and the instantaneous structure of wall-bounded turbulence \citep{hetsroni1994heat,Meinders1999, Gurka2004,Antonia1988,Abe2004,Abe2009,kim2020prediction}. Early experimental work with point measurements of heat transfer fluctuations reports a remarkable similarity between the conditionally sampled Reynolds heat flux and Reynolds shear stress \citep{perry1976experimental}. In particular, a quadrant analysis revealed that sweep and burst events are accompanied, with high probability, by equivalent events in the heat flux. \citet{iritani1985heat} visualized this by showing that the turbulent temperature field near the wall is similar to the flow field in the near-wall region, with the aid of temperature-sensitive liquid crystals and hydrogen bubbles to visualize thermal streaks and velocity streaks, respectively. \citet{kong2000direct} confirmed this strong correlation between the existence of both thermal and velocity streaks using \ac{DNS} of TBL flow. The spanwise spacing of the velocity and thermal streaks was estimated by means of two-point correlations, and equaled $100$ viscous lengths for isothermal boundary conditions. The viscous length scale, $l^*$, equals $\nu/{u_\tau}$, with $\nu$ being the fluid kinematic viscosity and ${u_\tau} \equiv {\sqrt{{\tau_w} / \rho}}$ being the mean friction velocity (where $\rho$ is the fluid density). Larger spacings (of approximately $150 l^*$) were observed for the iso-heat-flux boundary condition, due to the breakup of equation similarity between heat and momentum. The \ac{DNS} by \citet{Abe2001} analyzed turbulent heat transfer in channel flows at various values of the friction Reynolds number, $Re_\tau \equiv u_\tau \delta/\nu$, up to a value of $640$ (here $\delta$ is the boundary layer thickness). According to \citet{Abe2001}, the spanwise correlation of the wall temperature fields includes a negative peak at low Reynolds numbers, but this peak diminishes with an increase in $Re_\tau$ and is smaller than 5\% at $Re_\tau=640$ (thus suggesting that the wall temperature field becomes more broadband). Later results by \citet{Abe2004} confirmed this trend, providing results up to $Re_\tau = 1020$. Abe et al. ascribed the lack of negative spanwise-coherence in the wall-temperature field to clustering of the near-wall streaks in higher Reynolds number flows.  Recent simulations were performed over an even wider range of $Re_\tau$ (reaching values of $5\,000$), and for various Prandtl numbers and thermal boundary conditions \citep{pirozzoli2016passive,alcantara2021direct,alcantara2021directjfm}. There it was found that the intensity of thermal fluctuations increases with $Re_\tau$. 

Given the strong relation between wall-shear stress fluctuations (and off-the-wall turbulence quantities) and the convective heat transfer coefficient, $h$, it is evident that the latter can act as a surrogate for non-intrusive input data for flow estimation. \citet{kim2020prediction} recently utilized the strong relation between $\tau_w$ and $h$ for predictions of turbulent heat transfer; further experimental evidence by \citet{Miozzi2024} revolved around inferring the skin friction vector field around a wall-mounted cube, from time-resolved temperature maps of a heated wall obtained using a functional coating of temperature-sensitive paint. Hence, the experimentally-acquired convective heat transfer coefficient is informative of the off-the-wall velocity state.

For time-resolved measurements this is also true when utilizing the currently available high-sensitivity and high-repetition-rate IR cameras. In this regard, the instantaneous heat transfer fields have been measured in air by \citet{nakamura2013} and \citet{RAIOLA2017}, following the seminal work by \citet{hetsroni1994heat}.
Synchronized measurements were also obtained with \ac{IR} thermography and \ac{PIV} in water by \citet{gurka2004detecting} and \citet{foroozan2023}. However, when concentrating on air flows at moderate-to-high Reynolds numbers, the higher frequencies involved and the relatively small thermal conductivity of air challenge the accurate measurements of convective heat transfer in turbulent air flows. Recent progress has been made through data analysis, such as the procedure developed by \citet{cuellar2024measuring} to ensure a measurement uncertainty of the fluctuating Nusselt number $Nu$ below 10\%. 

The measurement of time-resolved convective heat transfer with a heated thin foil requires detailed measurements of wall-temperature fluctuations, which are the result of the fluctuating heat transfer coefficient and the foil's thermal inertia. Generally speaking, a thin foil can be used for this purpose if its Fourier number (the ratio between the flow characteristic time and the time scale of heat diffusion through the foil thickness) is much larger than one \cite{astarita2012infrared}. However, the amplitude of temperature fluctuations is attenuated by the foil thermal inertia, leading to small values which are difficult to measure when high frequencies are involved. Literature studies discussed above are either performed at low Reynolds numbers ($Re_\theta \lesssim 1000$) or in water where the flow velocity is lower and thermal conductivity is higher, leading to larger temperature fluctuations for a given Reynolds number.

In the present work, we explore 1) whether it is possible to measure for an air flow the fluctuating convective heat transfer coefficient in a \ac{TBL} with sufficient temporal resolution to capture meaningful turbulent events, and 2) how the measured spatiotemporal fields scale with Reynolds number. To this purpose, we employ high-repetition-rate \ac{IR} thermography, coupled with a heated-thin-foil heat transfer sensor beneath a zero-pressure-gradient (ZPG) TBL at two Reynolds numbers of $Re_\tau \approx 990$ and $Re_\tau \approx 1800$. Following a description of the experimental setup and the methodology in section \ref{sec:Setup}, we discuss the spatiotemporal maps of the heat transfer coefficient in section \ref{sec:Wall_results}, and statistical two-point correlations and energy spectra in \ref{sec:Spatial_analysis}. A frequency-wavenumber spectral analysis is carried out in section \ref{sec:Conv_Vel} to examine the convective velocities of different streamwise scales in the heat transfer fields.

%% file: Sections/2-Experimental_Setup.tex
\section{Experimental Setup}
\label{sec:Setup}

\subsection{Flow setup and statistics}

The experimental campaign was conducted in an open-loop wind tunnel facility within the Faculty of Aerospace Engineering, at the Delft University of Technology. A short overview of the main features of the facility is provided in the foregoing. A more detailed description can be found in Refs.~ \citep{Dacome2024,Baars_Dacome_Lee_2024}. 

The wind tunnel has a contraction ratio of $4.7:1$ and can attain a maximum velocity of approximately $16.5\,\text{m/s}$. The flow is accelerated through a contraction with an exit cross-sectional area of $0.60 \times 0.60\,\text{m}^2$. Two test sections, each $1.80\,\text{m}$ in length with polycarbonate walls for optical access, were placed downstream of the contraction exit plane. A new boundary layer developed on the bottom wall, which was suspended in the flow with the aid of a knife-edge design forming a bleed for the inflow boundary layer. For tripping the boundary layer, P40 sandpaper was used on all four walls at the inlet of the upstream test section. The ceiling of the test sections was adjusted to establish a streamwise, ZPG development of the \ac{TBL} over the bottom wall with a total length of $3.60\,\text{m}$.

\begin{figure}[t]
    \raggedright \hspace{5mm} (a) \hspace{69mm} (b)\\
    \centering
    \includegraphics[width=\textwidth]{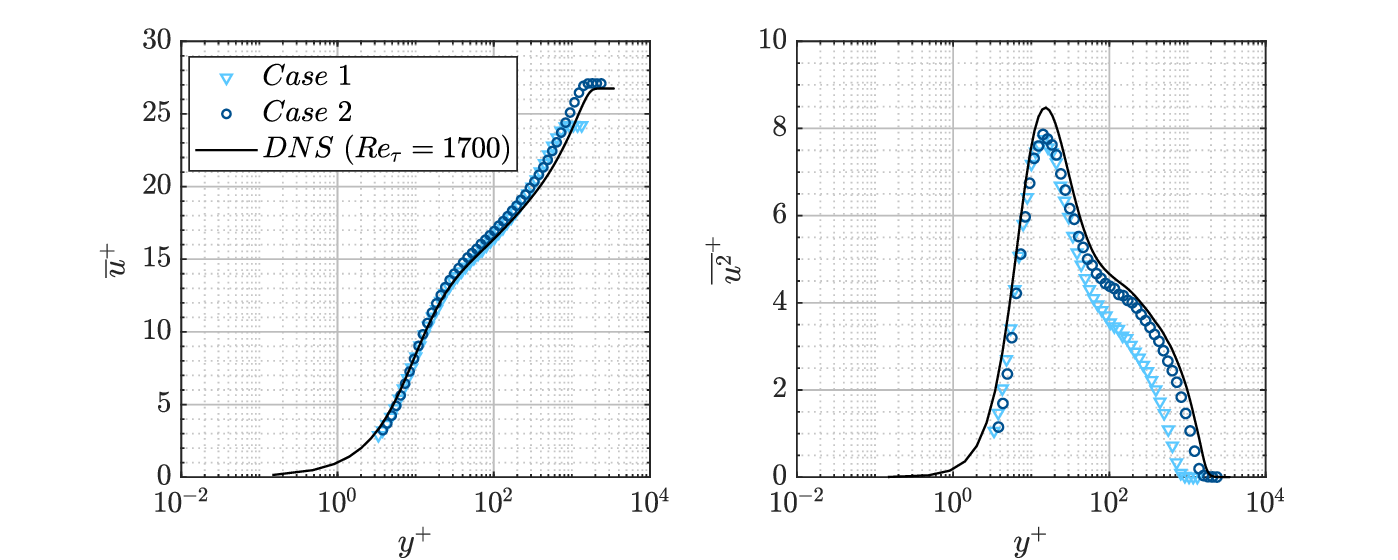}
    \caption{Wall-normal profiles of (a) the streamwise mean velocity, and (b) the streamwise turbulence kinetic energy, based on hot-wire data for Case 1 and Case 2 \citep{Dacome2024}. Experimental data are compared to profiles from DNS of TBL flow \citep{Sillero2014}.}
    \label{fig:TBL}
\end{figure}

A modular panel centered at approximately $x^\prime = 3.00$\,\text{m} ($x^\prime = 0$ coincides with the downstream edge of the sandpaper trip) accommodated the wall-embedded heated-thin-foil sensor, as shown in Fig.~\ref{fig:Setup_schem}(a). Experiments were performed at two distinct free-stream velocities of $5\,\text{m/s}$ and $10\,\text{m/s}$, and resulted in two nominal friction Reynolds numbers of $Re_\tau \approx 990$ and $Re_\tau \approx 1800$. These two testing conditions are further referred to as ``Case 1" and ``Case 2", respectively. Figure~\ref{fig:TBL} presents profiles of both the streamwise mean velocity and the streamwise velocity variance, measured with hot-wire anemometry. Boundary layer parameters were inferred by fitting the mean velocity profile to a composite profile with log-law constants of $\kappa = 0.384$ and $B = 4.17$ \citep{chauhan:2009a}. Table~\ref{tab:Table_TBL_char} reports the boundary layer parameters at the measurement location. Here, $\theta$ is the momentum thickness and the viscous length and time scales are denoted with symbols $l^*$ and $t^*$, respectively.

\begin{table}[t]
\begin{center}
    \captionsetup{singlelinecheck=false}
    \caption{Experimental parameters of the TBL flow at the two testing conditions and at the nominal streamwise location for measurements (where the heated thin-foil sensor was placed).}
    \input{Tables/TABLE_TBL}
    \label{tab:Table_TBL_char}
\end{center}
\end{table}

\subsection{Convective heat transfer measurement with Infrared Thermography}

\begin{figure}
    \raggedright (a) \hspace{37mm} (b) \hspace{43mm} (c) \\
    \centering
    \includegraphics[width=\textwidth]{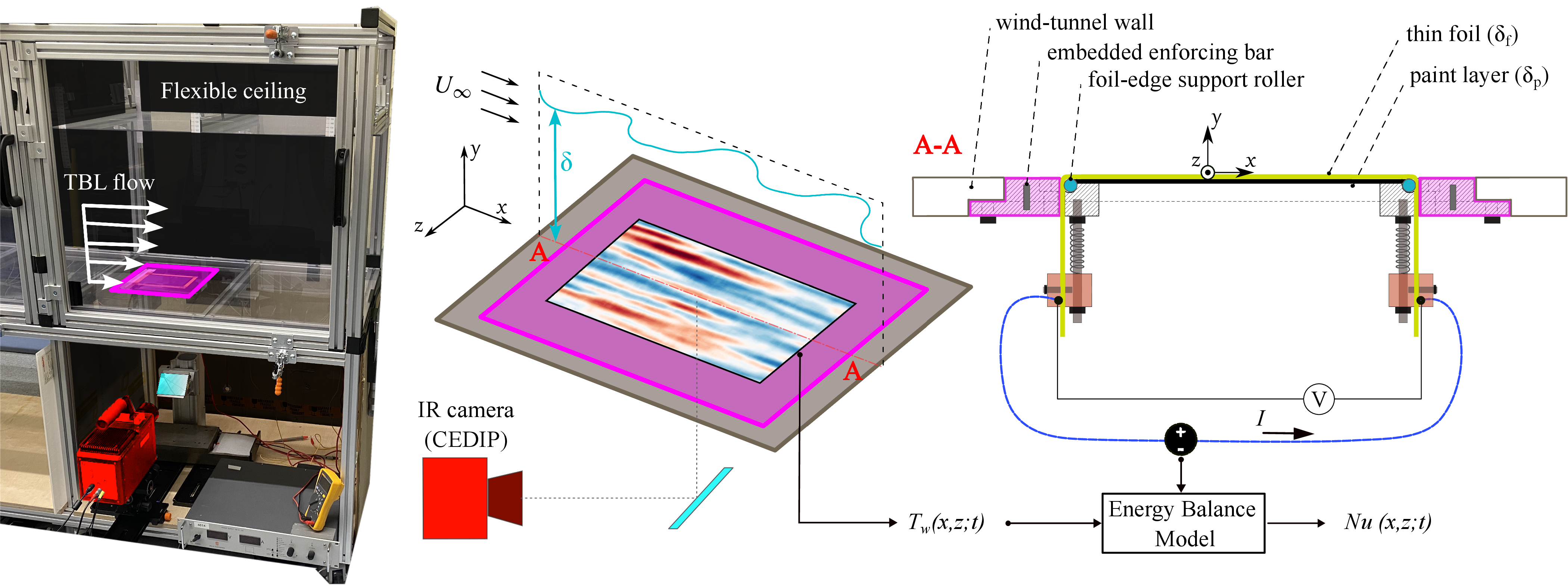}
    \caption{(a) Photograph of the experimental setup embedded in the wind tunnel facility at Delft University of Technology. The setup includes an \ac{IR} thermography system and a heated thin-foil sensor. (b) Schematic of the measurement area, and (c) cross-sectional view of the heated-thin-foil sensor.}
    \label{fig:Setup_schem}
\end{figure}

Convective heat transfer fluctuations on the wall were measured using high-repetition-rate \ac{IR} thermography, combined with a flush-mounted heated thin-foil sensor \citep{NAKAMURA2009}. Schematics of the setup with this heated-thin-foil sensor are shown in Figs.~\ref{fig:Setup_schem}(b) and \ref{fig:Setup_schem}(c). The setup included an \ac{IR}-transparent optical path for capturing temperature fluctuations on the rear-side of the heated foil with an \ac{IR} camera. A high-sensitivity IR camera was employed (CEDIP Titanium 530L model), comprising a frame rate of up to $20\,\text{kHz}$. Its Mercury Cadmium Telluride detector has a resolution of $320 \times 256$ pixels and has a noise equivalent temperature difference (NETD) of approximately $25\,\text{mK}$.

The heated-thin-foil sensor consisted of a $10\,\mu\text{m}$ thick stainless-steel foil, heated by a uniformly applied direct current across its leading- and trailing-edges relative to the flow. To enhance temperature measurements, the foil was coated with a thin layer of high-emissivity black paint. The foil was classified as thermally thin based on the estimated Biot number:
\begin{equation}
  Bi = h\left(\frac{\delta_{f}}{k_f} + \frac{\delta_{p}}{k_p}\right) < 0.1, 
  \label{eq:biot}
\end{equation}

\noindent where $h$ is the convective heat transfer coefficient, $\delta_{f}$ is the thickness of the foil, $k_f$ is the thermal conductivity of the foil, and $\delta_{p}$ and $k_p$ denote the same properties but now for the pain layer.  This layer of black matte high-emissivity paint ($\epsilon = 0.95 \pm 0.02$) was applied to minimize the measurement uncertainty of temperature fluctuations. A Biot number $Bi < 0.1$ implies that the thermal resistance within the material is negligible. The temperature distribution across the foil's thickness can thus be assumed to be uniform. In this study, the $Bi$ number is of the order of $10^{-4}$.

Moreover, to ensure that the foil is at a constant temperature through its thickness, also referring to the unsteady temperature variations, the modified Fourier number should be much larger than one \cite{astarita2012infrared}, i.e. the characteristic time of conduction through the foil should be much smaller than the characteristic time of the problem. By taking into account both the foil and the paint layers, the modified Fourier number can be defined as: 
\begin{equation}
 Fo = \frac{t_c}{\frac{\rho_f c_f  \delta_f^2}{k_f}+\frac{\rho_p c_p  \delta_p^2}{k_p}},  
  \label{eq:fourier}
\end{equation}

\noindent
where $\rho_f$ and $\rho_p$ are foil and paint densities, $c_f$ and $c_p$ are foil and paint heat capacities, and $t_c$ is the characteristic time of the problem which can be expressed for instance in outer units as the ratio of the boundary layer thickness and the free-stream velocity, $\delta/U_\infty$. According to this modified definition, the calculated values for $Fo$ in Case 1 and Case 2 are $7.7$ and $3.8$, respectively. This ensures that the characteristic time of the sensor is smaller than the characteristic time of the problem. The foil has a temporal response fast enough to track the typical heat transfer fluctuations in the flow if $Fo>>1$ \cite{astarita2012infrared}, otherwise the resulting temperature fluctuations might be attenuated at the higher frequencies. The importance of the paint contribution in equation \ref{eq:fourier}, lowering the value of the Fourier number, suggests the need for improved techniques for foil coating in future works. 

As illustrated in the cross-sectional view of Fig.~\ref{fig:Setup_schem}(c), the foil was securely stretched between a pair of copper clamps where the voltage was applied. The clamps are fitted with compressed springs to adjust the foil tension. The heated area of the foil is calculated based on the stretched length of the foil between these two clamps. The total length and width of the foil are $233\,\text{mm}$ and $103\,\text{mm}$, respectively, giving a heated area of $2.4 \times 10^{-2}\,\text{m}^2$. Note that this area is different from the measurement area which is set by the cutout in the sensor plate. This sensor plate was reinforced with two embedded metal bars to prevent bending due to the spring tension. To reduce contact resistance and local heating, the electrical connection between the copper and foil was achieved using $1\,\text{mm}$ thick Indium wires placed in a triangular engraving on the surface of the copper bars. Additionally, two roller bars supported the edges normal to the flow direction to ensure the foil remained stretched without any deformations. A $120 \times 80 \,\text{mm}$ cutout section was removed from the sensor plate as the sensing area to allow the \ac{IR} camera to view the foil from the non-flow exposed side. The acquisition parameters of this experiment along with their corresponding values are presented in Table~\ref{tab:Table_acq}. 

\begin{table*}
    \centering
    \captionsetup{singlelinecheck=false}
    \caption{Acquisition parameters.}
    \begin{threeparttable}
	\input{Tables/TABLE_acquisition}
    \end{threeparttable}
    \label{tab:Table_acq}
\end{table*}

Raw measurements of temperature fluctuations require processing to yield fields of $Nu$ fluctuations. For this, the convective heat transfer coefficient $h$ of a heated foil with surface temperature $T_w$, surrounded by a fluid at adiabatic wall temperature $T_{aw}$, can be estimated through an energy balance, as shown in Eq.~\eqref{eq:heat_balance}. The parameters related to the heated-thin-foil sensor, along with their corresponding values and uncertainties, are presented in Table \ref{tab:Table_parameters}.

\begin{equation}
    h = \frac{\dot{q}_{in} - \dot{q}_{cd} - \dot{q}_{rd} - B}{T_w - T_{aw}}
    \label{eq:heat_balance}
\end{equation}

In Eq.~\eqref{eq:heat_balance}, $\dot{q}_{in}$ is the input heat flux due to Joule heating, $\dot{q}_{cd}$ is the conduction heat flux through the foil, $\dot{q}_{rd}$ is the radiation heat flux, and $B$ is the thermal inertia term. The input heat flux $\dot{q}_{in}$ is obtained from the Joule heating of the foil, as:
\begin{equation}
 \dot{q}_{in} = {IV}/{A},
\end{equation}
where $I$ is the current applied through the foil, $V$ is the voltage differential between the copper clamps holding the foil, and $A$ is the surface area of the thin foil stretched between the two pairs of copper clamps. The radiation heat flux is estimated using the Stefan-Boltzmann law, considering the surface emissivity $\epsilon$ of the paint layer covering the foil and assuming that the whole wind tunnel laboratory can be modeled as a blackbody at ambient temperature $T_{amb}$: 
\begin{equation}
\dot{q}_{rd} = \sigma \epsilon (T_w^4 - T_{amb}^4),
\end{equation}
where $\sigma$ is the Stefan-Boltzmann constant.
The lateral conduction heat flux $\dot{q}_{cd}$ is obtained from the Laplacian of the wall temperature, multiplied by the in-plane thermal conductance:
\begin{equation}
\dot{q}_{cd} = -(k_f \delta_f + k_p \delta_p) \nabla^2 T_w,
\label{eq:lateral_conduction}
\end{equation}
where the subscripts $f$ and $p$ refer to foil and paint, respectively. Here, the computation of second spatial derivatives is possible thanks to the high spatial resolution ($1.31\,\text{pixels/mm}$) of temperature measurements, although it is challenged by the measurement noise, which requires spatial filtering \citep{RAIOLA2017,cuellar2024measuring}. 

Lastly, term $B$ in Eq.~\eqref{eq:heat_balance} represents the heat flux absorbed or released by the foil due to its heat capacity. It is computed as follows:
\begin{equation}
    B=\left( c_{f}\rho_f\delta_f + c_{p}\rho_p\delta_p\right)\left(\frac{\partial{T_w}}{\partial{t}}\right).
    \label{Eq_5}
\end{equation}

The temporal derivative was calculated using a central difference scheme with the frame interval as the time step. The \ac{IR} camera operates at a frame rate of $300\,\text{Hz}$ for Case 1 (low $Re_\tau$) and $500\,\text{Hz}$ for Case 2 (higher $Re_\tau$). This leads to similar non-dimensional acquisition frequencies when expressed in outer units (4.356 and 3.565 for Case 1 and Case 2, respectively, as outlined in Table \ref{tab:Table_acq}). However, since the viscous length decreases for increasing $Re_\tau$, this leads to a much lower frame rate when non-dimensionalized in inner units. As shown in the results section, given the large streamwise length of the wall patterns measured, the selected acquisition frequencies are deemed sufficient to capture the flow dynamics.

\begin{table*}
    \captionsetup{singlelinecheck=false}
    \caption{Parameters and their values used in the energy balance model of the sensor.}
    \label{tab:Table_parameters}
    \centering
    \begin{threeparttable}
        \input{Tables/TABLE_parameters}
        \begin{tablenotes}
            \item[1] Each of the two individual values correspond to Case 1 and Case 2 in the experiment.
        \end{tablenotes}
    \end{threeparttable}
\end{table*}

Fine-scale noise in the spatial-temporal distribution significantly affects the calculation of the space and time derivative terms in Eqs.~\eqref{eq:lateral_conduction} and~\eqref{Eq_5}. Additionally, a slow convection cell may form on the foil side that is not exposed to the flow. To eliminate these erroneous effects, raw data were filtered with a process similar to the one applied by \citet{cuellar2024measuring}. A high-pass filter removes features with a characteristic time of at least $250$ times longer than one eddy turnover time ($\delta/U_\infty$). 
Additionally, following \citet{RAIOLA2017}, a feature-based filtering, based on \ac{POD} is employed to remove any residual measurement noise; the number of modes retained was determined using the elbow method \citep{Cattell1966}. This procedure is needed to recover information about low-amplitude/high-frequency heat transfer fluctuations which, although measured by the sensor, may lead to temperature fluctuations smaller than the NETD of the IR camera.

Finally, the Nusselt number was calculated from the convective heat transfer coefficient:
\begin{equation}
    Nu = \frac{h \cdot \delta}{k_a},
    \label{eq:nusselt}
\end{equation}
with $\delta$ being set as the characteristic length of the problem, and $k_a$ being the thermal conductivity of air.
Measurement uncertainty of $Nu$, estimated following the method of \citet{MOFFAT19883} and employing the uncertainties reported in Table~\ref{tab:Table_parameters}, was found to be lower than $11\%$.

Analysis of the frequency spectrum in the experimental datasets revealed a consistent peak at \(39 \, \text{Hz}\) (\(\pm 2 \, \text{Hz}\)). This peak is likely caused by an external source, hypothesized to be the internal cooling system of the IR camera used during the experiments. This issue was previously observed on another IR camera by \citet{meola2015measurements}. To eliminate this disturbance, a band-pass filter targeting the range $37 \, \text{Hz}$ to $41 \, \text{Hz}$ was applied on the time-varying spatially averaged temperature across all datasets. This approach was performed on local patches of the image to take into account spatial non-uniformness due to the Narcissus effect \citep{howard1982narcissus} of the IR camera sensor. 

%% file: Tables/TABLE_TBL.tex
\begin{tabular}{c |c |c |c |c |c |c |c |c }
    \toprule
    Case  & $U_\infty (m/s)$  & $\delta (mm)$ & $\theta (mm)$  & $Re_\theta$   & $Re_\tau$ &   $u_\tau (m/s)$  &  $l^*\equiv \nu/u_\tau (\mu m)$  &  $t^*=\nu/u_\tau^2 (\mu s)$  \\[5pt]
    \midrule
     1     & 5    &  72.6    &   6.89   &   2248   &   988   &   0.204  & 73.5 & 360.2 \\[5pt]
     2     & 10   &  71.3    &   6.86   &   4572   &   1783  &   0.364  & 40.0 & 109.9 \\[5pt]
    \bottomrule
\end{tabular}

%% file: Tables/TABLE_acquisition.tex


\scriptsize 
\begin{tabular}{p{2.7cm}|p{1.1cm}p{1.3cm}p{1.3cm}|p{0.9cm}p{1.3cm}p{1.3cm}|p{0.9cm}p{1.3cm}p{1.3cm}}
\toprule
\textbf{Parameter}       & \multicolumn{3}{c|}{\textbf{Physical Units}}            & \multicolumn{3}{c|}{\textbf{Inner Units}}    & \multicolumn{3}{c}{\textbf{Outer Units}} \\
                         \cmidrule(r){2-4}                                         \cmidrule(l){5-7}                                        \cmidrule(l){8-10}
                         & \textbf{} & \textbf{Case 1}   & \textbf{Case 2}     & \textbf{}         & \textbf{Case 1}         & \textbf{Case 2}         & \textbf{}         & \textbf{Case 1}         & \textbf{Case 2}         \\
\midrule
Acquisition rate $f_s$          & Hz            & 300               & 500                 & $\frac{f_sl^*}{u_\tau}$                   & 0.108                   & 0.055                   & $\frac{f_s\delta}{U_\infty}$                   & 4.356                   & 3.565                   \\
Acquisition time $T_s$        & s             & 60                & 56                  & $\frac{T_su_\tau}{l^*}$                    & \(1.67 \times 10^5\)    & \(5.10 \times 10^5\)    & $\frac{T_sU_\infty}{\delta}$                    & \(4.13 \times 10^3\)    & \(7.85 \times 10^3\)    \\
Window width $\Delta z$            & m             & 0.08              & 0.08                & $\frac{\Delta z}{l^*}$                    & 1088.4                  & 2000.0                 & $\frac{\Delta z}{\delta}$                    & 1.10                    & 1.12                    \\
Window length $\Delta x$            & m             & 0.12              & 0.12                & $\frac{\Delta x}{l^*}$                    & 1632.7                  & 3000.0                 & $\frac{\Delta x}{\delta}$                    & 1.65                    & 1.68                    \\
Image resolution         & px/mm         & 1.31              & 1.31                & px$/l^*$                & 0.0963                  & 0.0524                 &  px$/\delta$               & 95.106                  & 93.403                  \\
\bottomrule
\end{tabular}

%% file: Tables/TABLE_parameters.tex
\begin{tabular}{ p{6.5cm} p{2cm} p{3.5cm} p{2cm}}
    \toprule
     Quantity                       & Symbol      & Value                       & Units             \\[5pt]
    \midrule
   Electrical current             &$I$          &14.5,17.5$\pm0.01$\tnote{1}  & $A$             \\[3pt]
     Electrical voltage             &$V$          &3.2,\,3.8$\pm0.01$\tnote{1}  & $V$             \\[3pt]
     Heated foil area               &$A$          &0.024$\pm2.5\times 10^{-5}$  & $m^2$           \\[3pt]
     Ambient temperature            &$T_{amb}$    &24.7                         & $^{\circ}C$       \\[3pt]
     Thermal conductivity (foil)    &$k_f$        &16.3 $\pm2$                  & $W/(m K)$       \\[3pt]
     Thermal conductivity (paint)   &$k_p$        &1.4 $\pm0.1$                 & $W/(m K)$       \\[3pt]
     Thickness (foil)               &$\delta_f$   &10 $\pm0.1$                  & $\mu m$         \\[3pt]
     Thickness (paint)              &$\delta_p$   &20 $\pm0.1$                  & $\mu m$         \\[3pt]
     Heat capacity (foil)           &$c_{p,f}$    &500$\pm5$                    & $J/(kg K)$      \\[3pt]
     Heat capacity (paint)          &$c_{p,p}$    &5000$\pm50$                  & $J/(kg K)$      \\[3pt]
     Density (foil)                 &$\rho_f$     &7930$\pm50$                  & $kg/m^3$        \\[3pt]
     Density (paint)                &$\rho_p$     &1300$\pm50$                  & $kg/m^3$        \\[3pt]
     Emissivity (foil)              &$\epsilon_f$   &0.17$\pm0.02$                & $-$               \\[3pt]
     Emissivity (paint)              &$\epsilon_p$   &0.95$\pm0.02$                & $-$               \\[3pt]
     Stefan-Boltzmann constant      &$\sigma$     &$5.67\times 10^{-8}$         & $W/(m^2 K^4)$   \\[3pt]
     \bottomrule
\end{tabular}

%% file: Sections/3-Wall_Measurement.tex
\section{Results and discussion}
\label{sec:Results}

Results are organized into three subsections. First, samples of the instantaneous temperature and convective heat transfer maps are presented, referred to as the flow \textit{footprint}. Secondly, second-order spatial statistics, comprising two-point correlations and two-dimensional spectral analysis, are covered to quantify the streamwise and spanwise scales of the turbulence present within the footprint. Subsequently, a third subsection focuses on an assessment of the convective velocities associated with the broadband range of turbulence structures affecting the footprint. Literature has shown that the convective velocity is strongly scale-dependent \citep{Liu_Gayme_2020}, especially for scales that are most energetic in the near-wall region.

%% file: Sections/3.1.tex
\subsection {Understanding the footprint}
\label{sec:Wall_results}

\begin{figure}
        \raggedright (a) \hspace{75mm} (b)
        \includegraphics[width=0.49\textwidth]{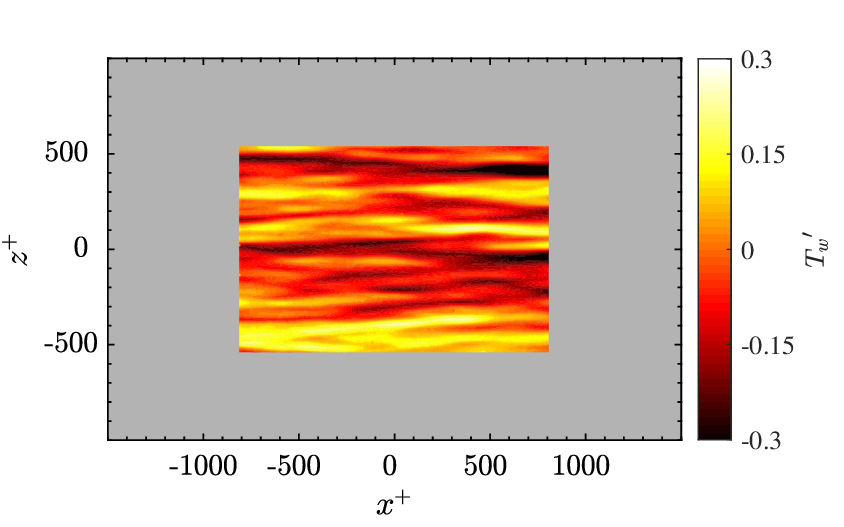} 
        \includegraphics[width=0.49\textwidth]{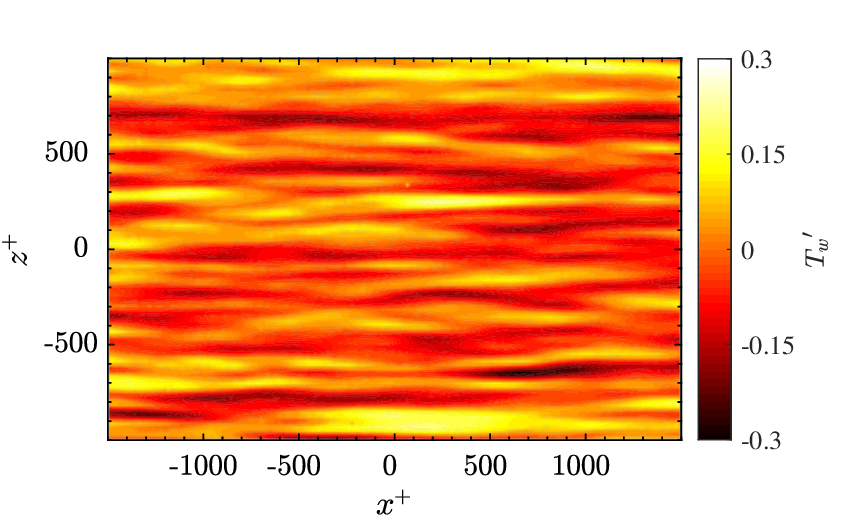}\\[-10pt]
        \raggedright (c) \hspace{75mm} (d)
        \includegraphics[width=0.49\textwidth]{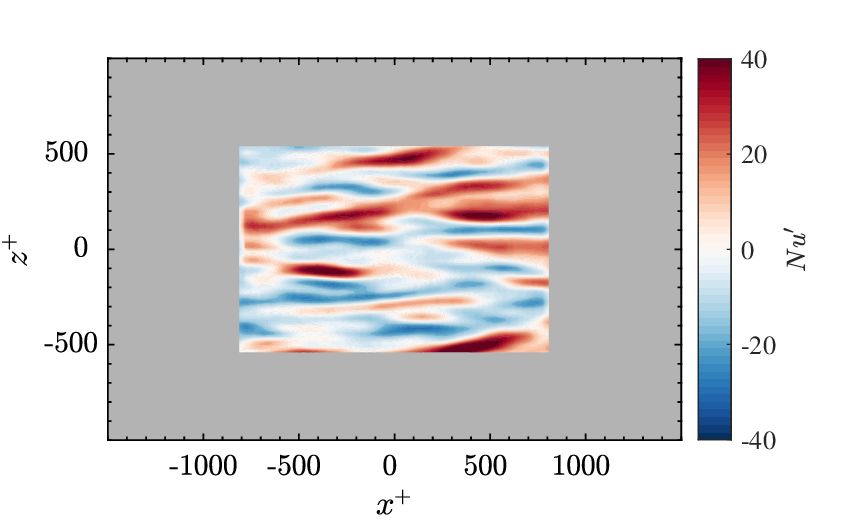} 
        \includegraphics[width=0.49\textwidth]{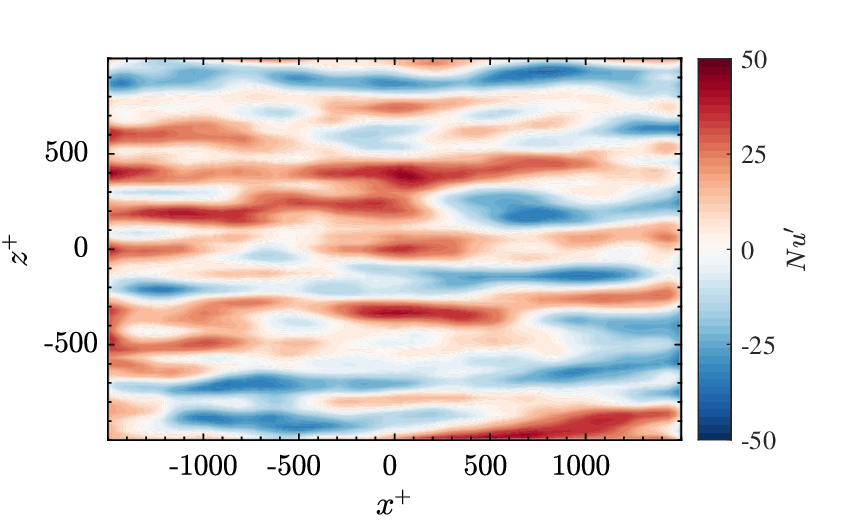} 
    \caption{(a,b) Instantaneous maps of the wall-temperature, $T_w$, and (c,d) Nusselt number, $Nu$, for both Case 1 (left) and Case 2 (right).}
    \label{fig:temp_Nu_maps}
\end{figure}

Figure~\ref{fig:temp_Nu_maps} reports contour plots of sample fields of the instantaneous fluctuating wall temperature, $T_w'=T_w-\overline{T_w}$, and Nusselt number $Nu'=Nu-\overline{Nu}$. The average values, indicated with an overline, are obtained with spatial and temporal averaging of the raw $T_w(x,z,t)$ and $Nu(x,z,t)$ fields. The origin of the coordinate system used in this study is located at the streamwise/spanwise center of the heat transfer sensor. Axes of all sub-plots in Fig.~\ref{fig:temp_Nu_maps} are expressed in inner units and comprise the same range. As such, the fixed sensor size is larger in terms of viscous units for the higher Reynolds-number data of Case 2. 

Qualitatively, regions of positive temperature fluctuations correspond generally to areas of integrated reduced convective heat transfer (leading to an increase of temperature of the foil over time), while negative temperature fluctuation regions indicate enhanced convective heat transfer. It has to be remarked though that due to the unsteady term in Eq.~\eqref{Eq_5} this correspondence might not be obvious (while in slower flows this is more evident, see \citep{foroozan2023}). To quantify spatial variations in temperature, the mean and standard deviation of $T_w-T_{aw}$ were calculated across the measurement area. The mean wall-temperature increase for the lower $Re$ was found to be $\overline{T_w-T_{aw}} = 55.28\,\text{K}$, with a standard deviation of $0.12\,\text{K}$, and for the higher $Re_\tau$ was found to be $\overline{T_w-T_{aw}} = 53.77 \,\text{K}$, with a standard deviation of $0.091\,\text{K}$. While these values of heating may seem large, it must be remarked that the induced buoyancy effects are negligible, being the Richardson number smaller than $10^-2$ for the lower $Re_\tau$ case. The lower level of temperature fluctuations for the case at higher $Re_\tau$, suggests an attenuation due to the foil thermal inertia at higher frequencies, although of limited intensity.

The $Nu'$ maps in Figs.~\ref{fig:temp_Nu_maps}(c) and \ref{fig:temp_Nu_maps}(d) highlight regions of enhanced or reduced instantaneous heat transfer intensity. High values correspond to strong convective heat transfer; by assuming an analogy between $Nu$ and the skin friction coefficient $c_f$, these regions are subject to higher streamwise velocity. Synchronized velocity and heat transfer measurements are needed to verify this analogy and to quantify the phase delay between flow structures and wall heat transfer patterns. The mean Nusselt number for the lower $Re_\tau$ was calculated as $\overline{Nu} = 68 \pm 5$, with a standard deviation of $16.2$, and for the higher $Re_\tau$ was calculated as $\overline{Nu} = 106.5 \pm 10$, with a standard deviation of $17$, reflecting the variability in convective heat transfer due to the TBL. Average values of the Nusselt number are in reasonably good agreement with empirical relations available from the open literature for Reynolds numbers based on outer scaling. For instance, \citet{mcadams1926heat} reports a correlation for fully-developed turbulent pipe flows (where the Reynolds number $Re_D$ is based on the pipe diameter), $\overline{Nu} = 0.023 Re_D^{0.8} Pr^{0.4}$, of which its range of validity is $\quad 0.7 \leq Pr \leq 120$ and $\quad Re_D \geq 10^4$. 
According to this correlation, the expected values for $\overline{Nu}$ in Case 1 and Case 2 are $66$ and $113$, respectively. 
For what concerns the $Nu$ fluctuation intensity the results by \citet{Abe2004} ---who report wall-temperature fluctuations--- suggest that they should increase proportionally to the average Nusselt number. The small increase in the $Nu$ standard deviation observed in Case 2 if compared to Case 1 suggests thus an attenuation of the high-frequency features due to the lower Fourier number of the sensor (Eq.~\ref{eq:fourier}) in the higher-$Re_\tau$ case.

When inspecting the pattern of temperature fluctuations, the presence of elongated streaks is seen. This result is in agreement with the well-known presence of streaks populating the near-wall region of the boundary layer. In the $Nu'$ contour maps, streaks exhibit a spanwise spacing of approximately $150l^*$ and $300l^*$, at the lower and higher Reynolds numbers, respectively. Our present visualization of streaky patterns in the wall-fields enriches a long-standing trend in the literature on near-wall flow visualization of elongated streaks \citep{Kline1967,Smith_Metzler_1983,Klewicki1995,Smits2011}. The streaks demonstrate a wide range of lengths, some are found to exceed $1000l^*$ at both Reynolds numbers. The spanwise spacing, larger than the classical $\mathcal{O}(100l^*)$ of velocity streaks is in agreement with the finding by \citet{kong2000direct} for iso-heat-flux boundary conditions and with those of Abe \emph{et al.} \citep{Abe2001,Abe2004} claiming larger streak width for increasing Reynolds number.

%% file: Sections/3.2.tex
\subsection{Time-averaged spatial statistics of the footprint}
\label{sec:Spatial_analysis}
\begin{figure}
    \raggedright (a) \hspace{75mm}(b)
    \includegraphics[width=0.49\textwidth]{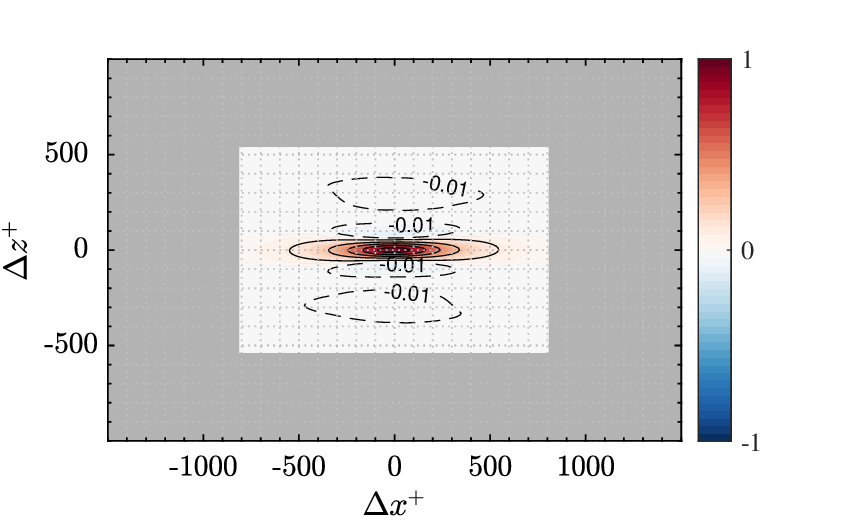} 
    \includegraphics[width=0.49\textwidth]{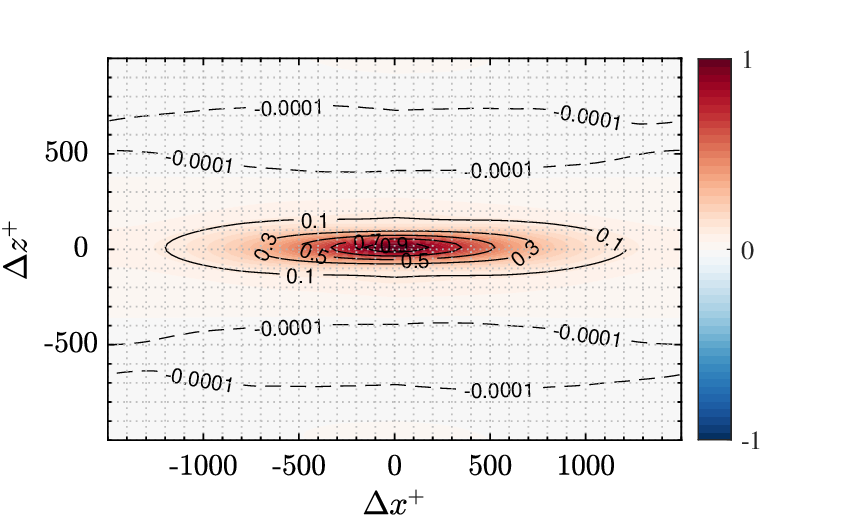} 
    \caption{Maps of the two-point correlation coefficient, $\rho_{NuNu}$, as defined in Eq.~\eqref{eq:3.1} for (a) Case 1, and (b) Case 2. Positive contour levels are from $0.1$ to $0.9$ with increments of $0.2$. One iso-contour corresponding to a negative value is also shown.}
    \label{fig:temp_aurocorr_map}
\end{figure}

To statistically quantify the main features and lengthscales of the patterns in the $Nu'$ fields of Fig.~\ref{fig:temp_aurocorr_map}, second-order statistics are considered in terms of two-point correlations and two-dimensional, spatial spectral analysis. Two-point correlations are considered following the normalized correlation coefficient, defined as:
\begin{equation}\label{eq:3.1}
    \rho_{NuNu}(\Delta x,\Delta z) = \frac{\langle Nu^\prime(x,z,t) Nu^\prime(x + \Delta x,z + \Delta z,t)\rangle_t}{\overline{Nu^{\prime2}}}, 
\end{equation}
where $\langle \dots \rangle_t$ indicates ensemble-averaging in time and the denominator is the $Nu$ variance evaluated based on the entire dataset (\emph{i.e.}, computed in space and time).

The two-point correlation map in Fig.~\ref{fig:temp_aurocorr_map} shows a streamwise-elongated arrangement, surrounded by shallow negative peaks. The presence of weak negative peaks surrounding the strong elongated positive peak is in agreement with the less prominent negative peaks with increasing Reynolds number discussed in the literature \citep[\emph{e.g.},][]{Abe2001, Abe2004, li2009dns}. The elongated correlation region shown for the lower $Re$, becomes longer and thicker for the higher $Re$; in this case, the negative values for the correlation map within the observed region become smaller, confirming the trend expected by \citet{Abe2001} for increasing $Re_\tau$. 

\begin{figure}
\vspace{5mm}
    \raggedright (a) \hspace{75mm}(b)\\
    \centering
    \includegraphics[width=0.49\textwidth]{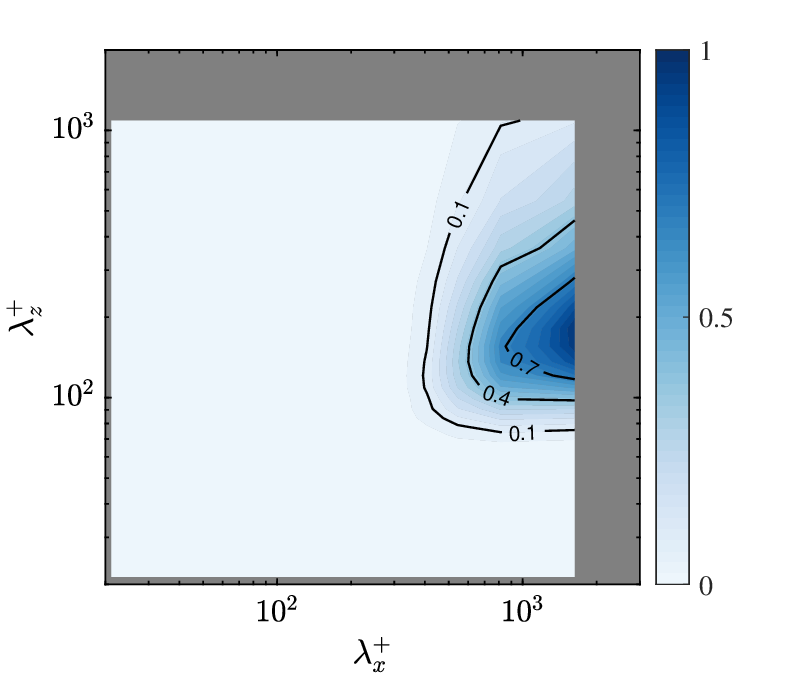} 
    \includegraphics[width=0.49\textwidth]{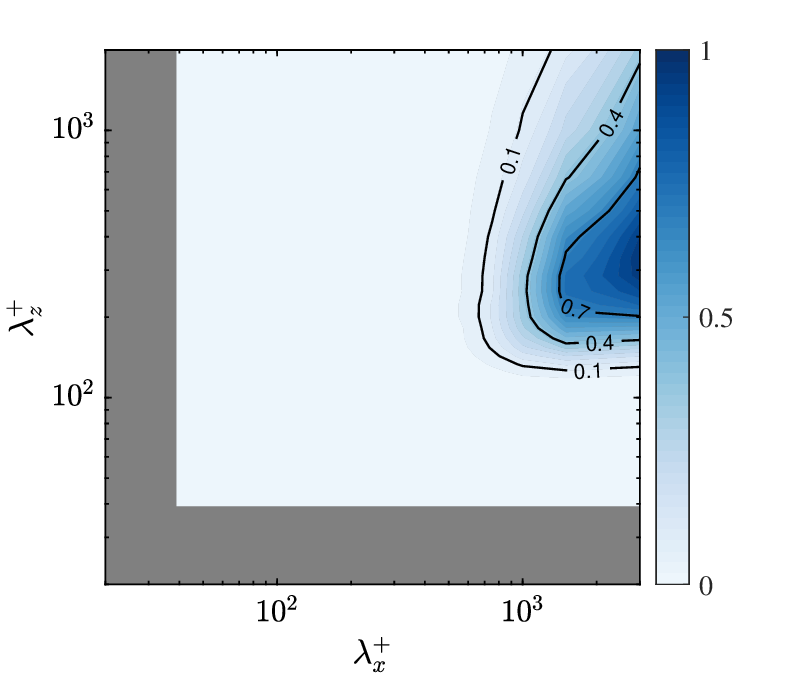} 
    \caption{Pre-multiplied two-dimensional energy spectra of Nusselt number fluctuations, $k_x^+ k_z^+ \phi_{NuNu}$, for (a) Case 1, and (b) Case 2. The contour levels contain $10\%$, $40\%$ and $70\%$ of the maximum power-spectral density.}
    \label{fig:power_spectrum_Re}
\end{figure}

The two-dimensional spatial spectrum is computed to assess the scale distribution in space. Here the spectrum is taken according to:
\begin{equation}\label{eq:spatialspec}
    \phi_{NuNu}(\lambda_x,\lambda_z) = \left\langle \widetilde{Nu}\left(\lambda_x,\lambda_z;t\right)\right\rangle_t,
\end{equation}
with the tilde indicating the two-dimensional Fourier transform in both spatial directions, \emph{e.g.}, $\widetilde{Nu}\left(\lambda_x,\lambda_z;t\right) = \mathcal{F}\left[Nu(x,z,t)\right]$. Two-dimensional spectra are shown in Fig.~\ref{fig:power_spectrum_Re} for both Case 1 and Case 2. Again, the sub-plots comprise axes scaled with inner units, spanning a similar range, and thus the data cover different regions due to the fixed physical sensor size and resolution of the \ac{IR} camera.

At the lower Reynolds number, most energy in the 2D spectrum resides at a spanwise wavelength of approximately $\lambda_z^+ = 150$. Such a wavelength is larger than the typical $100$ wall unit-width of near-wall streaks and is instead in agreement with wavelengths observed for the temperature fluctuations further from the wall. For instance, the spatial spectra of temperature fluctuations for $Pr=1$ reported by \citet{balasubramanian2023direct}, at a lower Reynolds number and at a wall-normal distance of $y^+=30$, have a similar spectral distribution. This might hint that the measured $Nu$ fields are the footprint of several hierarchies of coherent flow features, spanning different sizes and wall-normal extends. Corroborating this argument, for the larger Reynolds number the dominant spanwise wavelength is $300$ to $350$ wall units. This increase of the spanwise wavelength with increasing Reynolds number is consistent with the results reported by \citet{li2009dns}. 

When focusing on the energy distribution as a function of the streamwise wavelength, the dominant region of energy resides at wavelengths larger than $\lambda_x^+ = 1500$. At the higher Reynolds number, the dominant wavelength shifts to even higher wavelengths beyond the size of the sensor ($\lambda_x^+ > 3000$). This shift reflects the larger scale separation at higher Reynolds numbers. However, it is important to note that the spatial resolution and the dimensions of the sensors used in the \ac{IR} thermography system can impose certain limitations on the ability to effectively capture and analyze small structures.

Overall, the results presented in the foregoing demonstrate the ability of the non-intrusive heated-thin-foil sensor to capture the (mean) convective heat-transfer coefficient in a TBL air flow, as well as the spatial structure of streamwise-elongated streaks in the footprint of the TBL flow. When an inner-normalization is used, this structure is not invariant with $Re_\tau$. This is due to a combination of an attenuation of the smaller-scale structures in the higher $Re_\tau$ scenario (Case 2, comprising less sensor-resolution in viscous units), as well as the presence of more large-scale velocity fluctuations in the TBL flow and consequently in the measured footprint.


%% file: Sections/3.3.tex
\subsection{Convective velocity estimation}
\label{sec:Conv_Vel}
\begin{figure}
    \centering
    \raggedright \hspace{2mm} (a) \hspace{90mm}(b)
    \includegraphics[width=\textwidth]{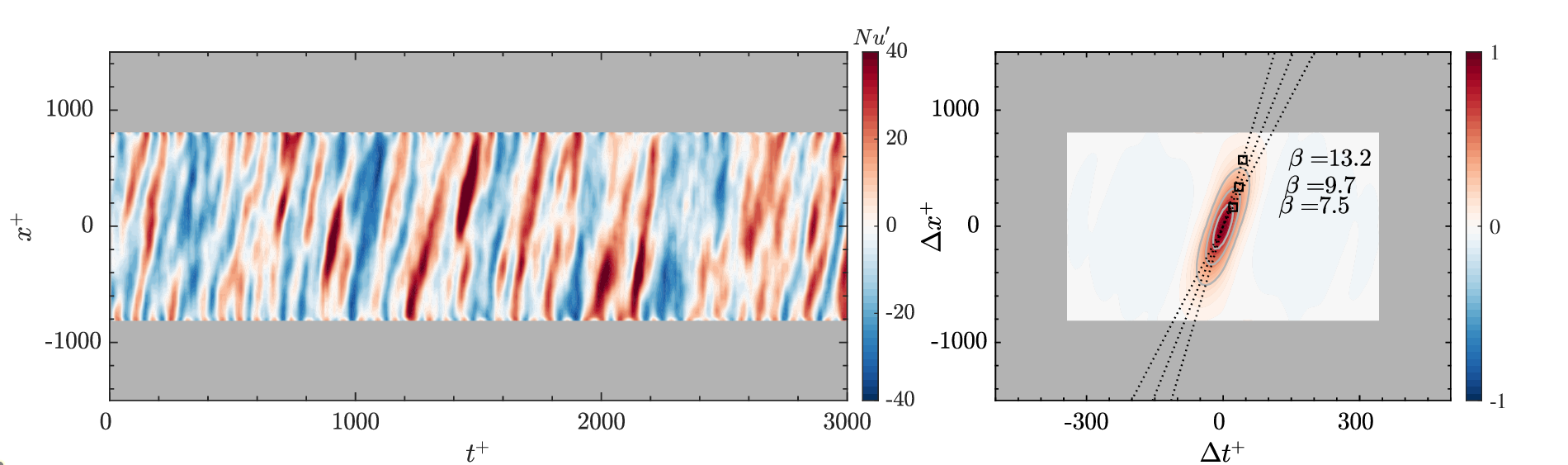}\\[-10pt]
    \raggedright \hspace{2mm} (c) \hspace{90mm}(d)
    \includegraphics[width=\textwidth]{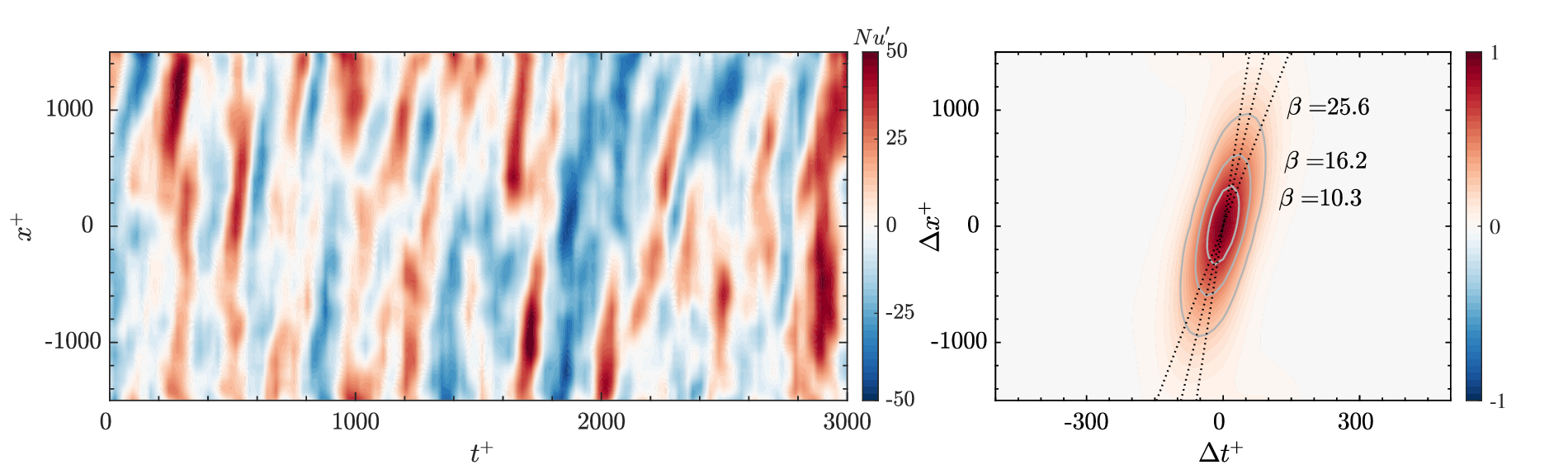} 
    \caption{(a,c) Space-time contour maps of $Nu$ fluctuations at the mid-span of the sensor ($z = 0$), and (b,d) maps of the two-point correlation coefficient, $\rho_{NuNu}$, for both Case 1 (top row) and Case 2 (bottom row). Within the two-point correlation maps, contour levels are drawn at values of 0.25, 0.50 and 0.75; their inclinations are characterized by the $\beta$ values and suggest that flow structures of a different size comprise different convection velocities.}
    \label{fig:Space-time_Autocorr}
\end{figure}

Given the time-resolution of the measurement (recall Table~\ref{tab:Table_acq}), the temporal evaluation of the spatial footprint can be assessed to study how the flow-induced pattern of the footprint convects downstream. A raw visualization of the convective nature is shown in Fig.~\ref{fig:Space-time_Autocorr}, in which the space-time maps of $Nu'$ are shown over a short time interval (fixed range in terms of $t^+$) for both Case 1 (Fig.~\ref{fig:Space-time_Autocorr}a) and Case 2 (Fig.~\ref{fig:Space-time_Autocorr}c). These space-time contours are shown for the mid-span of the $Nu$ field, so for $Nu(x^+,z=0,t^+)$. Persistent streaky structures are observed over the domain. The spatiotemporal evolution of the features observed in the convective heat transfer maps depicts the convective velocity of the flow structures' footprint at the wall.

In the literature, there are several methods to determine the convection velocity, $U_c$, of turbulent structures, \citep[\emph{e.g.},][]{wills_1964,Choi1990,Kim1993,Jeon1999,Abe2001}. Using space-time separations, the convective velocity of coherent structures can be estimated by identifying their convective length and time, after which the convective velocity is simply taken as $U_c = {\Delta x}/{\Delta t}$. This approach assumes that turbulent eddies are convected downstream without significant deformation over short time (or length) scales. The applicability of this method relying on Taylor's hypothesis has been extensively reviewed in the literature \citep{DELÁLAMO_JIMÉNEZ_2009,MOIN_2009}. 

The space-time correlations can be generated from the space-time data to obtain preliminary estimates of $U_c$ values. This follows Eq.~\eqref{eq:3.1}, but with a temporal offset instead of a spanwise spatial one, to yield $\rho_{NuNu}(\Delta x,\Delta t)$. Space-time correlations corresponding to Cases 1 and 2 are shown in Figs.~\ref{fig:Space-time_Autocorr}(b) and ~\ref{fig:Space-time_Autocorr}(d), respectively. The inclined iso-contours of the two-point correlation maps reveal that the convection velocities vary for different turbulent length scales, and that the convection velocity is thus scale-dependent (broadband). And even though all scale-dependent information is masked due to various scales contributing to the correlation for a certain offset ($\Delta x$ and $\Delta t$), the two-point correlation maps can still aid in assessing the general trend of the convective nature, as has been well established with similar approaches in turbulent flow relying on hot-wire data \citep[\emph{e.g.},][]{ goldschmidt_young_ott_1981,wills_1964,Krogstad1998ConvectionVI}. 

The elongated contours in the streamwise direction suggest that large-scale turbulent structures maintain coherence over extended distances. 
In contrast, smaller structures, exhibit more compact contours, indicating lower convective velocities as they are convected at speeds closer to the local mean velocity near the wall. In Figs.~\ref{fig:Space-time_Autocorr}(b) and~\ref{fig:Space-time_Autocorr}(d), the $\beta$ values correspond to the space-time slope, following $\beta \equiv \Delta x^+/\Delta t^+$. These values thus reflect a convection velocity, $U_c^+$, and the orientation of the elongated contours across three scales (selected with the contour lines at $\rho_{NuNu}(\Delta x,\Delta t) = 0.25$, $0.5$ and $0.75$). A similar study is presented by \citet{Abe2004} for $Re_\tau=1020$, where \ac{DNS} of turbulent heat transfer is carried out in a channel flow.

The estimated convective velocities exhibit a distinct scale-dependent behavior. In the lower $Re_\tau$ case, the convection velocity of larger-scale structures is approximately $13u_\tau$, while smaller structures exhibited convection velocities in the range of $7 u_\tau$ to $8 u_\tau$. This trend aligns with previous findings in the literature, such as those by \citet{delalamo2004}, who reported similar scale-dependent velocities in \ac{DNS} of \acp{TBL}. They suggested that, while the convection velocity of the small scales is similar to the mean velocity above $y^+ \approx 10$ to $15$, the convection velocity of the large scales is close to the free-stream velocity even in the near-wall region and this results in a non-zero value for the averaged convection velocity at the wall. 
A comparison of the space–time correlations between two Reynolds numbers indicates that the inclination angle of the correlations $\beta$ for higher $Re_\tau$ is much steeper than that of the correlations for lower $Re_\tau$, suggesting a larger convection velocity.

\begin{figure}
    \raggedright (a) \hspace{75mm}(b)
    \includegraphics[width=0.49\textwidth]{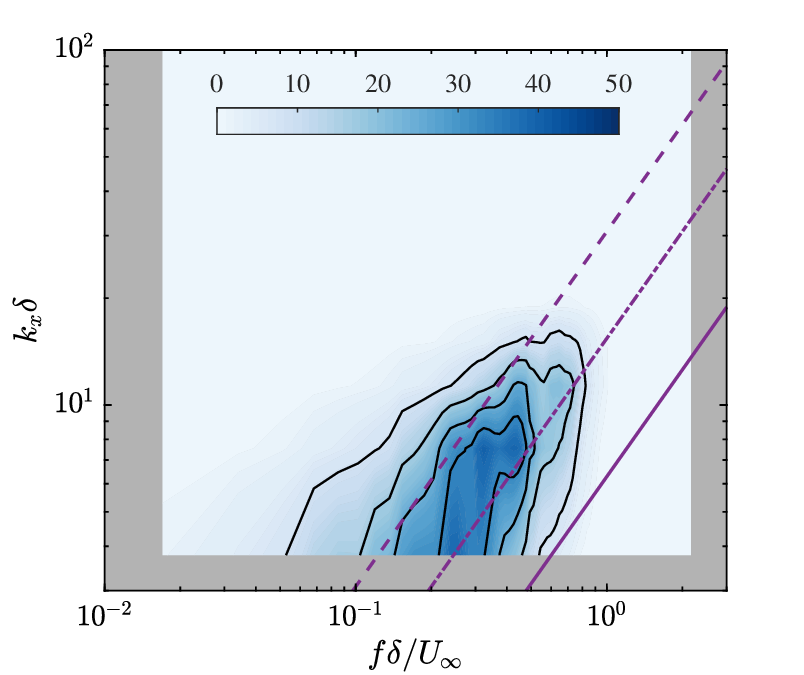}
    \includegraphics[width=0.49\textwidth]{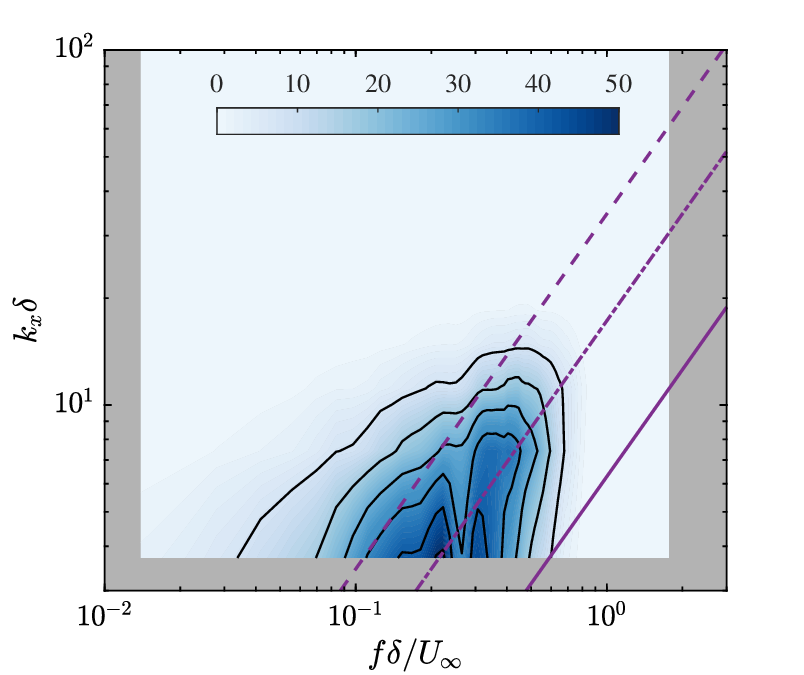}
    \caption{Pre-multiplied frequency-wavenumber spectra, $k_x^+ f^+ \phi_{NuNu}$, for (a) Case 1, and (b) Case 2. Trend lines are included to indicate constant convection velocities of $U_c = 5 u_\tau$ (\sampleline{dashed}), $U_c = 10 u_\tau$ (\sampleline{dash pattern=on .7em off .2em on .05em off .2em}), and $U_c = U_\infty$ (\sampleline{}).}
    \label{fig:freq-waveNo}
\end{figure}

It is possible to retain scale information by considering the space-time correlation in the wavenumber-frequency domain. For this we employ the wavenumber-frequency spectrum, taken from the data at a specific $z_i$ location as:
\begin{equation}\label{eq:spatialspec2}
    \phi_{NuNu}(\lambda_x,f) = \left\langle \widetilde{Nu}\left(\lambda_x,z_i;t\right)\right\rangle_z,
\end{equation}
with the tilde indicating here the two-dimensional Fourier transform along the space (in $x$) and time, \emph{e.g.}, $\widetilde{Nu}\left(\lambda_x,f\right) = \mathcal{F}\left[Nu(x,z_i,t)\right]$. Ensemble averaging is performed in time, and also by considering all $z_i$ locations. Frequency-wavenumber spectra are presented in pre-multiplied form in Fig.~\ref{fig:freq-waveNo}. Here, straight trend lines are added and are representative of various convection velocities, since $U_c = \frac{2\pi f}{k_x}$ with $f$ being the temporal frequency and $k_x$ being the spatial wavenumber in $x$. 


The observation of Fig.~\ref{fig:freq-waveNo} confirms the results of Fig.~\ref{fig:Space-time_Autocorr}: more elongated features (smaller wavenumber) travel faster, and convection velocities are larger at the larger Reynolds number. This allows us to claim that the features observed in the $Nu'$ maps measured at the wall are the footprint of wall-attached features populating the boundary layer. Those who extend further from the wall are also more elongated and convent at a larger velocity.

\begin{figure}
    \centering
    \includegraphics[width=0.55\textwidth]{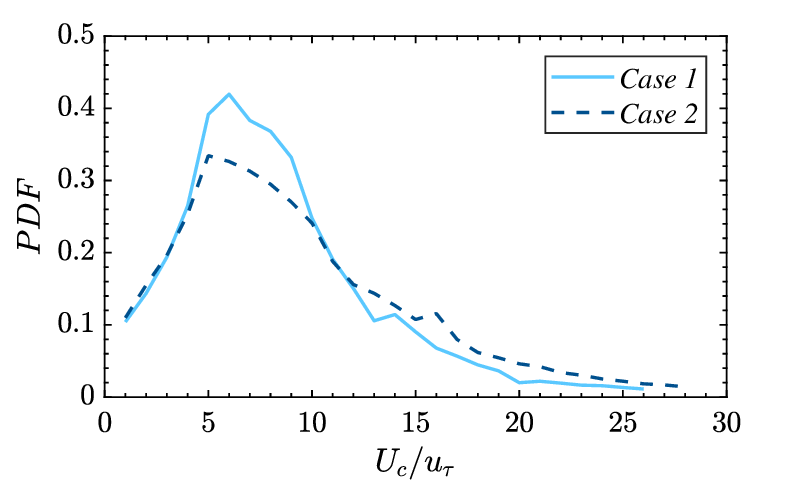}
    \caption{Probability density functions of the frequency-wavenumber spectra, in bands of $U_c/u_\tau \pm1$.}
    \label{fig:Uc-energy}
\end{figure}

According to \citet{wills_1964} the $f,k_x$-spectrum can be used to obtain another measure of convective velocity that is a function of the wavenumber. Figure~\ref{fig:Uc-energy} complements this analysis by presenting the probability density function (PDF) of \(f,k_x\) distributions for velocity bands normalized by \( u_\tau \) (\( U_c/u_\tau \pm 1 \)). The PDF function is calculated as:

\begin{equation} \label{eq:3.2}
    \text{PDF} = \frac{\sum_{k_x/f = m - r_1}^{m + r_2} \phi_{NuNu}(k_x,f)}{\sum_{}^{} \phi_{NuNu}(k_x,f)} \cdot (r_1 + r_2)
\end{equation}

\noindent
where $(r_1 + r_2)$ is the range length. Through this approach, we can highlight the contribution to the total energy content from scales moving at certain convection velocities. From Fig.~\ref{fig:Uc-energy}, we can infer that the bulk of the energy moves at convection velocities in the range \( 5u_\tau \) to \( 15u_\tau \). And so, most of the turbulent activity contributing to the observed heat transfer at the wall originates from structures within this velocity range. From the velocity profiles in Fig.~\ref{fig:TBL}, we can hypothesize that such structures are predominantly located below $y^+ = 50$. These findings are consistent with the interpretation that the dominant energy in the wall-normal direction originates from the near-wall region, where smaller turbulent structures contribute significantly to the convective heat transfer. At the higher Reynolds number the amount of flow structures convected at larger velocities is greater, confirming the trends reflected in Figs.~\ref{fig:Space-time_Autocorr} and \ref{fig:freq-waveNo}.

%% file: Sections/4-Conclusions.tex
\section{Conclusions}
\label{sec:conclusion}

This study demonstrates the potential of non-intrusive sensing techniques for characterizing the TBL footprint through the analysis of wall-temperature fluctuations. Using a high-repetition-rate IR thermography system, combined with a heated-thin-foil sensor embedded within the wall, we successfully captured the spatiotemporal dynamics of convective heat transfer beneath a zero-pressure-gradient turbulent boundary layer at two Reynolds numbers, \(Re_\tau \approx 990\) and \(Re_\tau \approx 1800\).

The results reveal that the Nusselt number fluctuations serve as a thermal footprint of turbulent boundary layer structures. At both Reynolds numbers, streamwise-elongated streaks were observed with dominant spanwise wavelengths of approximately $150$ wall units for \(Re_\tau \approx 1000\) and $300$ wall units for \(Re_\tau \approx 1800\). 
These streaks represent not only near-wall structures populating the near-wall region, but also the footprint of larger, wall-attached features, confirming the increasing role of large-scale turbulence with higher Reynolds numbers. 

A two-dimensional spectral analysis of the $Nu$ fields identified distinct turbulent scales. At higher Reynolds numbers, dominant streamwise wavelengths exceeded $3000$ wall units, indicating an increased separation of scales and highlighting the influence of large-scale structures on wall-bounded heat transfer fluctuations. The spatial resolution of the system adequately captured these patterns, although finer-scale features remain attenuated due to sensor limitations.

Convective velocity estimation further elucidated the footprint at the wall of boundary layer coherent structures. Large-scale structures were found to convect at speeds close to the free-stream velocity (\(U_\infty\)), while smaller-scale features exhibited lower velocities, consistent with the local mean flow. Namely, at $Re_\tau = 1800$, the convective velocity of large structures was approximately $0.8 U_\infty$, while smaller structures exhibited velocities ranging from $0.4 U_\infty$ to $0.6 U_\infty$. 

Although the current resolution adequately captures the dominant turbulent features present in the flow, there remains a significant opportunity for enhancement, especially considering the attenuation of high-frequency, fine-scale turbulence. Future work should focus on improving sensor resolution to capture finer-scale turbulence. By addressing these challenges, non-intrusive sensing could become an even more powerful tool for turbulence analysis and real-time control.

In conclusion, this study demonstrates the effectiveness of non-intrusive sensing techniques in capturing the complex dynamics of boundary layer turbulence and heat transfer. The findings offer valuable insights into the spatial and temporal scales of turbulent structures, their convective velocities, and their impact on convective heat transfer. These findings are paving the way to targeted flow control strategies, as they prove we can identify flow scales from the footprint of the convective heat transfer.

%% file: MAIN.bbl
\begin{thebibliography}{62}%
\makeatletter
\providecommand \@ifxundefined [1]{%
 \@ifx{#1\undefined}
}%
\providecommand \@ifnum [1]{%
 \ifnum #1\expandafter \@firstoftwo
 \else \expandafter \@secondoftwo
 \fi
}%
\providecommand \@ifx [1]{%
 \ifx #1\expandafter \@firstoftwo
 \else \expandafter \@secondoftwo
 \fi
}%
\providecommand \natexlab [1]{#1}%
\providecommand \enquote  [1]{``#1''}%
\providecommand \bibnamefont  [1]{#1}%
\providecommand \bibfnamefont [1]{#1}%
\providecommand \citenamefont [1]{#1}%
\providecommand \href@noop [0]{\@secondoftwo}%
\providecommand \href [0]{\begingroup \@sanitize@url \@href}%
\providecommand \@href[1]{\@@startlink{#1}\@@href}%
\providecommand \@@href[1]{\endgroup#1\@@endlink}%
\providecommand \@sanitize@url [0]{\catcode `\\12\catcode `\$12\catcode `\&12\catcode `\#12\catcode `\^12\catcode `\_12\catcode `\%12\relax}%
\providecommand \@@startlink[1]{}%
\providecommand \@@endlink[0]{}%
\providecommand \url  [0]{\begingroup\@sanitize@url \@url }%
\providecommand \@url [1]{\endgroup\@href {#1}{\urlprefix }}%
\providecommand \urlprefix  [0]{URL }%
\providecommand \Eprint [0]{\href }%
\providecommand \doibase [0]{https://doi.org/}%
\providecommand \selectlanguage [0]{\@gobble}%
\providecommand \bibinfo  [0]{\@secondoftwo}%
\providecommand \bibfield  [0]{\@secondoftwo}%
\providecommand \translation [1]{[#1]}%
\providecommand \BibitemOpen [0]{}%
\providecommand \bibitemStop [0]{}%
\providecommand \bibitemNoStop [0]{.\EOS\space}%
\providecommand \EOS [0]{\spacefactor3000\relax}%
\providecommand \BibitemShut  [1]{\csname bibitem#1\endcsname}%
\let\auto@bib@innerbib\@empty
\bibitem [{\citenamefont {Cattafesta}\ and\ \citenamefont {Sheplak}(2011)}]{cattafesta2011actuators}%
  \BibitemOpen
  \bibfield  {author} {\bibinfo {author} {\bibfnamefont {L.~N.}\ \bibnamefont {Cattafesta}}\ and\ \bibinfo {author} {\bibfnamefont {M.}~\bibnamefont {Sheplak}},\ }\bibfield  {title} {\bibinfo {title} {Actuators for active flow control},\ }\href@noop {} {\bibfield  {journal} {\bibinfo  {journal} {Annu. Rev. Fluid Mech.}\ }\textbf {\bibinfo {volume} {43}},\ \bibinfo {pages} {247} (\bibinfo {year} {2011})}\BibitemShut {NoStop}%
\bibitem [{\citenamefont {Astarita}\ and\ \citenamefont {Carlomagno}(2012)}]{astarita2012infrared}%
  \BibitemOpen
  \bibfield  {author} {\bibinfo {author} {\bibfnamefont {T.}~\bibnamefont {Astarita}}\ and\ \bibinfo {author} {\bibfnamefont {G.~M.}\ \bibnamefont {Carlomagno}},\ }\href@noop {} {\emph {\bibinfo {title} {Infrared thermography for thermo-fluid-dynamics}}}\ (\bibinfo  {publisher} {Springer Science \& Business Media},\ \bibinfo {year} {2012})\BibitemShut {NoStop}%
\bibitem [{\citenamefont {G{\"u}emes}\ \emph {et~al.}(2019)\citenamefont {G{\"u}emes}, \citenamefont {Discetti},\ and\ \citenamefont {Ianiro}}]{guemes2019sensing}%
  \BibitemOpen
  \bibfield  {author} {\bibinfo {author} {\bibfnamefont {A.}~\bibnamefont {G{\"u}emes}}, \bibinfo {author} {\bibfnamefont {S.}~\bibnamefont {Discetti}},\ and\ \bibinfo {author} {\bibfnamefont {A.}~\bibnamefont {Ianiro}},\ }\bibfield  {title} {\bibinfo {title} {Sensing the turbulent large-scale motions with their wall signature},\ }\href@noop {} {\bibfield  {journal} {\bibinfo  {journal} {Phys. Fluids}\ }\textbf {\bibinfo {volume} {31}} (\bibinfo {year} {2019})}\BibitemShut {NoStop}%
\bibitem [{\citenamefont {Encinar}\ and\ \citenamefont {Jim{\'e}nez}(2019)}]{encinar2019logarithmic}%
  \BibitemOpen
  \bibfield  {author} {\bibinfo {author} {\bibfnamefont {M.~P.}\ \bibnamefont {Encinar}}\ and\ \bibinfo {author} {\bibfnamefont {J.}~\bibnamefont {Jim{\'e}nez}},\ }\bibfield  {title} {\bibinfo {title} {Logarithmic-layer turbulence: a view from the wall},\ }\href@noop {} {\bibfield  {journal} {\bibinfo  {journal} {Phys. Rev. Fluids}\ }\textbf {\bibinfo {volume} {4}},\ \bibinfo {pages} {114603} (\bibinfo {year} {2019})}\BibitemShut {NoStop}%
\bibitem [{\citenamefont {G{\"u}emes}\ \emph {et~al.}(2021)\citenamefont {G{\"u}emes}, \citenamefont {Discetti}, \citenamefont {Ianiro}, \citenamefont {Sirmacek}, \citenamefont {Azizpour},\ and\ \citenamefont {Vinuesa}}]{guemes2021coarse}%
  \BibitemOpen
  \bibfield  {author} {\bibinfo {author} {\bibfnamefont {A.}~\bibnamefont {G{\"u}emes}}, \bibinfo {author} {\bibfnamefont {S.}~\bibnamefont {Discetti}}, \bibinfo {author} {\bibfnamefont {A.}~\bibnamefont {Ianiro}}, \bibinfo {author} {\bibfnamefont {B.}~\bibnamefont {Sirmacek}}, \bibinfo {author} {\bibfnamefont {H.}~\bibnamefont {Azizpour}},\ and\ \bibinfo {author} {\bibfnamefont {R.}~\bibnamefont {Vinuesa}},\ }\bibfield  {title} {\bibinfo {title} {From coarse wall measurements to turbulent velocity fields through deep learning},\ }\href@noop {} {\bibfield  {journal} {\bibinfo  {journal} {Phys. Fluids}\ }\textbf {\bibinfo {volume} {33}} (\bibinfo {year} {2021})}\BibitemShut {NoStop}%
\bibitem [{\citenamefont {Guastoni}\ \emph {et~al.}(2021)\citenamefont {Guastoni}, \citenamefont {G{\"u}emes}, \citenamefont {Ianiro}, \citenamefont {D.}, \citenamefont {Schlatter}, \citenamefont {Azizpour},\ and\ \citenamefont {Vinuesa}}]{guastoni2021convolutional}%
  \BibitemOpen
  \bibfield  {author} {\bibinfo {author} {\bibfnamefont {L.}~\bibnamefont {Guastoni}}, \bibinfo {author} {\bibfnamefont {A.}~\bibnamefont {G{\"u}emes}}, \bibinfo {author} {\bibfnamefont {A.}~\bibnamefont {Ianiro}}, \bibinfo {author} {\bibfnamefont {S.}~\bibnamefont {D.}}, \bibinfo {author} {\bibfnamefont {P.}~\bibnamefont {Schlatter}}, \bibinfo {author} {\bibfnamefont {H.}~\bibnamefont {Azizpour}},\ and\ \bibinfo {author} {\bibfnamefont {R.}~\bibnamefont {Vinuesa}},\ }\bibfield  {title} {\bibinfo {title} {Convolutional-network models to predict wall-bounded turbulence from wall quantities},\ }\href@noop {} {\bibfield  {journal} {\bibinfo  {journal} {J. Fluid Mech.}\ }\textbf {\bibinfo {volume} {928}},\ \bibinfo {pages} {A27} (\bibinfo {year} {2021})}\BibitemShut {NoStop}%
\bibitem [{\citenamefont {Cu{\'e}llar}\ \emph {et~al.}(2024{\natexlab{a}})\citenamefont {Cu{\'e}llar}, \citenamefont {G{\"u}emes}, \citenamefont {Ianiro}, \citenamefont {Flores}, \citenamefont {Vinuesa},\ and\ \citenamefont {Discetti}}]{cuellar2024three}%
  \BibitemOpen
  \bibfield  {author} {\bibinfo {author} {\bibfnamefont {A.}~\bibnamefont {Cu{\'e}llar}}, \bibinfo {author} {\bibfnamefont {A.}~\bibnamefont {G{\"u}emes}}, \bibinfo {author} {\bibfnamefont {A.}~\bibnamefont {Ianiro}}, \bibinfo {author} {\bibfnamefont {{\'O}.}~\bibnamefont {Flores}}, \bibinfo {author} {\bibfnamefont {R.}~\bibnamefont {Vinuesa}},\ and\ \bibinfo {author} {\bibfnamefont {S.}~\bibnamefont {Discetti}},\ }\bibfield  {title} {\bibinfo {title} {Three-dimensional generative adversarial networks for turbulent flow estimation from wall measurements},\ }\href@noop {} {\bibfield  {journal} {\bibinfo  {journal} {J. Fluid Mech.}\ }\textbf {\bibinfo {volume} {991}},\ \bibinfo {pages} {A1} (\bibinfo {year} {2024}{\natexlab{a}})}\BibitemShut {NoStop}%
\bibitem [{\citenamefont {Cu{\'e}llar}\ \emph {et~al.}(2024{\natexlab{b}})\citenamefont {Cu{\'e}llar}, \citenamefont {Ianiro},\ and\ \citenamefont {Discetti}}]{cuellar2024some}%
  \BibitemOpen
  \bibfield  {author} {\bibinfo {author} {\bibfnamefont {A.}~\bibnamefont {Cu{\'e}llar}}, \bibinfo {author} {\bibfnamefont {A.}~\bibnamefont {Ianiro}},\ and\ \bibinfo {author} {\bibfnamefont {S.}~\bibnamefont {Discetti}},\ }\bibfield  {title} {\bibinfo {title} {{Some effects of limited wall-sensor availability on flow estimation with 3D-GANs}},\ }\href@noop {} {\bibfield  {journal} {\bibinfo  {journal} {Theor. Comput. Fluid Dyn.}\ ,\ \bibinfo {pages} {1}} (\bibinfo {year} {2024}{\natexlab{b}})}\BibitemShut {NoStop}%
\bibitem [{\citenamefont {Guastoni}\ \emph {et~al.}(2024)\citenamefont {Guastoni}, \citenamefont {Balasubramanian}, \citenamefont {Foroozan}, \citenamefont {G\"{u}emes}, \citenamefont {Ianiro}, \citenamefont {Discetti}, \citenamefont {Schlatter}, \citenamefont {Azizpour},\ and\ \citenamefont {Vinuesa}}]{guastoni2024fully}%
  \BibitemOpen
  \bibfield  {author} {\bibinfo {author} {\bibfnamefont {L.}~\bibnamefont {Guastoni}}, \bibinfo {author} {\bibfnamefont {A.~G.}\ \bibnamefont {Balasubramanian}}, \bibinfo {author} {\bibfnamefont {F.}~\bibnamefont {Foroozan}}, \bibinfo {author} {\bibfnamefont {A.}~\bibnamefont {G\"{u}emes}}, \bibinfo {author} {\bibfnamefont {A.}~\bibnamefont {Ianiro}}, \bibinfo {author} {\bibfnamefont {S.}~\bibnamefont {Discetti}}, \bibinfo {author} {\bibfnamefont {P.}~\bibnamefont {Schlatter}}, \bibinfo {author} {\bibfnamefont {H.}~\bibnamefont {Azizpour}},\ and\ \bibinfo {author} {\bibfnamefont {R.}~\bibnamefont {Vinuesa}},\ }\bibfield  {title} {\bibinfo {title} {Fully convolutional networks for velocity-field predictions based on the wall heat flux in turbulent boundary layers},\ }\bibfield  {journal} {\bibinfo  {journal} {Theor. Comput. Fluid Dyn.}\ }\textbf {\bibinfo {volume} {39}},\ \href {https://doi.org/10.1007/s00162-024-00732-y} {10.1007/s00162-024-00732-y} (\bibinfo {year} {2024})\BibitemShut {NoStop}%
\bibitem [{\citenamefont {Dacome}\ \emph {et~al.}(2024)\citenamefont {Dacome}, \citenamefont {M\"orsch}, \citenamefont {Kotsonis},\ and\ \citenamefont {Baars}}]{Dacome2024}%
  \BibitemOpen
  \bibfield  {author} {\bibinfo {author} {\bibfnamefont {G.}~\bibnamefont {Dacome}}, \bibinfo {author} {\bibfnamefont {R.}~\bibnamefont {M\"orsch}}, \bibinfo {author} {\bibfnamefont {M.}~\bibnamefont {Kotsonis}},\ and\ \bibinfo {author} {\bibfnamefont {W.~J.}\ \bibnamefont {Baars}},\ }\bibfield  {title} {\bibinfo {title} {Opposition flow control for reducing skin-friction drag of a turbulent boundary layer},\ }\href {https://doi.org/10.1103/PhysRevFluids.9.064602} {\bibfield  {journal} {\bibinfo  {journal} {Phys. Rev. Fluids}\ }\textbf {\bibinfo {volume} {9}},\ \bibinfo {pages} {064602} (\bibinfo {year} {2024})}\BibitemShut {NoStop}%
\bibitem [{\citenamefont {Kerherv{\'e}}\ \emph {et~al.}(2017)\citenamefont {Kerherv{\'e}}, \citenamefont {Roux},\ and\ \citenamefont {Mathis}}]{kerherve2017combining}%
  \BibitemOpen
  \bibfield  {author} {\bibinfo {author} {\bibfnamefont {F.}~\bibnamefont {Kerherv{\'e}}}, \bibinfo {author} {\bibfnamefont {S.}~\bibnamefont {Roux}},\ and\ \bibinfo {author} {\bibfnamefont {R.}~\bibnamefont {Mathis}},\ }\bibfield  {title} {\bibinfo {title} {Combining time-resolved multi-point and spatially-resolved measurements for the recovering of very-large-scale motions in high {R}eynolds number turbulent boundary layer},\ }\href@noop {} {\bibfield  {journal} {\bibinfo  {journal} {Exp. Therm. Fluid Sci.}\ }\textbf {\bibinfo {volume} {82}},\ \bibinfo {pages} {102} (\bibinfo {year} {2017})}\BibitemShut {NoStop}%
\bibitem [{\citenamefont {Discetti}\ \emph {et~al.}(2019)\citenamefont {Discetti}, \citenamefont {Bellani}, \citenamefont {{\"O}rl{\"u}}, \citenamefont {Serpieri}, \citenamefont {Sanmiguel~Vila}, \citenamefont {Raiola}, \citenamefont {Zheng}, \citenamefont {Mascotelli}, \citenamefont {Talamelli},\ and\ \citenamefont {Ianiro}}]{discetti2019characterization}%
  \BibitemOpen
  \bibfield  {author} {\bibinfo {author} {\bibfnamefont {S.}~\bibnamefont {Discetti}}, \bibinfo {author} {\bibfnamefont {G.}~\bibnamefont {Bellani}}, \bibinfo {author} {\bibfnamefont {R.}~\bibnamefont {{\"O}rl{\"u}}}, \bibinfo {author} {\bibfnamefont {J.}~\bibnamefont {Serpieri}}, \bibinfo {author} {\bibfnamefont {C.}~\bibnamefont {Sanmiguel~Vila}}, \bibinfo {author} {\bibfnamefont {M.}~\bibnamefont {Raiola}}, \bibinfo {author} {\bibfnamefont {X.}~\bibnamefont {Zheng}}, \bibinfo {author} {\bibfnamefont {L.}~\bibnamefont {Mascotelli}}, \bibinfo {author} {\bibfnamefont {A.}~\bibnamefont {Talamelli}},\ and\ \bibinfo {author} {\bibfnamefont {A.}~\bibnamefont {Ianiro}},\ }\bibfield  {title} {\bibinfo {title} {{Characterization of very-large-scale motions in high-Re pipe flows}},\ }\href@noop {} {\bibfield  {journal} {\bibinfo  {journal} {Exp. Therm. Fluid Sci.}\ }\textbf {\bibinfo {volume} {104}},\ \bibinfo {pages} {1} (\bibinfo {year} {2019})}\BibitemShut {NoStop}%
\bibitem [{\citenamefont {Pastuhoff}\ \emph {et~al.}(2013)\citenamefont {Pastuhoff}, \citenamefont {Yorita}, \citenamefont {Asai},\ and\ \citenamefont {Alfredsson}}]{pastuhoff2013enhancing}%
  \BibitemOpen
  \bibfield  {author} {\bibinfo {author} {\bibfnamefont {M.}~\bibnamefont {Pastuhoff}}, \bibinfo {author} {\bibfnamefont {D.}~\bibnamefont {Yorita}}, \bibinfo {author} {\bibfnamefont {K.}~\bibnamefont {Asai}},\ and\ \bibinfo {author} {\bibfnamefont {P.~H.}\ \bibnamefont {Alfredsson}},\ }\bibfield  {title} {\bibinfo {title} {Enhancing the signal-to-noise ratio of pressure sensitive paint data by singular value decomposition},\ }\href@noop {} {\bibfield  {journal} {\bibinfo  {journal} {Meas. Sci. Tech.}\ }\textbf {\bibinfo {volume} {24}},\ \bibinfo {pages} {075301} (\bibinfo {year} {2013})}\BibitemShut {NoStop}%
\bibitem [{\citenamefont {Gu}\ \emph {et~al.}(2024)\citenamefont {Gu}, \citenamefont {Discetti}, \citenamefont {Liu}, \citenamefont {Cao},\ and\ \citenamefont {Peng}}]{gu2024denoising}%
  \BibitemOpen
  \bibfield  {author} {\bibinfo {author} {\bibfnamefont {F.}~\bibnamefont {Gu}}, \bibinfo {author} {\bibfnamefont {S.}~\bibnamefont {Discetti}}, \bibinfo {author} {\bibfnamefont {Y.}~\bibnamefont {Liu}}, \bibinfo {author} {\bibfnamefont {Z.}~\bibnamefont {Cao}},\ and\ \bibinfo {author} {\bibfnamefont {D.}~\bibnamefont {Peng}},\ }\bibfield  {title} {\bibinfo {title} {Denoising image-based experimental data without clean targets based on deep autoencoders},\ }\href@noop {} {\bibfield  {journal} {\bibinfo  {journal} {Exp. Therm. Fluid Sci.}\ }\textbf {\bibinfo {volume} {156}},\ \bibinfo {pages} {111195} (\bibinfo {year} {2024})}\BibitemShut {NoStop}%
\bibitem [{\citenamefont {Amili}\ and\ \citenamefont {Soria}(2011)}]{amili2011film}%
  \BibitemOpen
  \bibfield  {author} {\bibinfo {author} {\bibfnamefont {O.}~\bibnamefont {Amili}}\ and\ \bibinfo {author} {\bibfnamefont {J.}~\bibnamefont {Soria}},\ }\bibfield  {title} {\bibinfo {title} {A film-based wall shear stress sensor for wall-bounded turbulent flows},\ }\href@noop {} {\bibfield  {journal} {\bibinfo  {journal} {Exp. Fluids}\ }\textbf {\bibinfo {volume} {51}},\ \bibinfo {pages} {137} (\bibinfo {year} {2011})}\BibitemShut {NoStop}%
\bibitem [{\citenamefont {Reynolds}(1874)}]{reynolds1874extent}%
  \BibitemOpen
  \bibfield  {author} {\bibinfo {author} {\bibfnamefont {O.}~\bibnamefont {Reynolds}},\ }\bibfield  {title} {\bibinfo {title} {On the extent and action of the heating surface of steam boilers},\ }in\ \href@noop {} {\emph {\bibinfo {booktitle} {Proceedings of the Manchester Literary and Philosophical Society}}},\ Vol.~\bibinfo {volume} {14}\ (\bibinfo {year} {1874})\ \bibinfo {note} {session 1874-1875}\BibitemShut {NoStop}%
\bibitem [{\citenamefont {Reynolds}(1961)}]{reynolds1961extent}%
  \BibitemOpen
  \bibfield  {author} {\bibinfo {author} {\bibfnamefont {O.}~\bibnamefont {Reynolds}},\ }\bibfield  {title} {\bibinfo {title} {On the extent and action of the heating surface of steam boilers},\ }\href@noop {} {\bibfield  {journal} {\bibinfo  {journal} {Int. J. Heat Mass Tran.}\ }\textbf {\bibinfo {volume} {3}},\ \bibinfo {pages} {163} (\bibinfo {year} {1961})}\BibitemShut {NoStop}%
\bibitem [{\citenamefont {Hetsroni}\ and\ \citenamefont {Rozenblit}(1994)}]{hetsroni1994heat}%
  \BibitemOpen
  \bibfield  {author} {\bibinfo {author} {\bibfnamefont {G.}~\bibnamefont {Hetsroni}}\ and\ \bibinfo {author} {\bibfnamefont {R.}~\bibnamefont {Rozenblit}},\ }\bibfield  {title} {\bibinfo {title} {Heat transfer to a liquid—solid mixture in a flume},\ }\href@noop {} {\bibfield  {journal} {\bibinfo  {journal} {Int. J. Multiphas. Fl.}\ }\textbf {\bibinfo {volume} {20}},\ \bibinfo {pages} {671} (\bibinfo {year} {1994})}\BibitemShut {NoStop}%
\bibitem [{\citenamefont {Meinders}\ \emph {et~al.}(1999)\citenamefont {Meinders}, \citenamefont {Hanjalic},\ and\ \citenamefont {Martinuzzi}}]{Meinders1999}%
  \BibitemOpen
  \bibfield  {author} {\bibinfo {author} {\bibfnamefont {E.~R.}\ \bibnamefont {Meinders}}, \bibinfo {author} {\bibfnamefont {K.}~\bibnamefont {Hanjalic}},\ and\ \bibinfo {author} {\bibfnamefont {R.~J.}\ \bibnamefont {Martinuzzi}},\ }\bibfield  {title} {\bibinfo {title} {Experimental study of the local convection heat transfer from a wall-mounted cube in turbulent channel flow},\ }\href {https://doi.org/10.1115/1.2826017} {\bibfield  {journal} {\bibinfo  {journal} {J. Heat Transf.}\ }\textbf {\bibinfo {volume} {121}},\ \bibinfo {pages} {564–573} (\bibinfo {year} {1999})}\BibitemShut {NoStop}%
\bibitem [{\citenamefont {Gurka}\ \emph {et~al.}(2004{\natexlab{a}})\citenamefont {Gurka}, \citenamefont {Liberzon},\ and\ \citenamefont {Hetsroni}}]{Gurka2004}%
  \BibitemOpen
  \bibfield  {author} {\bibinfo {author} {\bibfnamefont {R.}~\bibnamefont {Gurka}}, \bibinfo {author} {\bibfnamefont {A.}~\bibnamefont {Liberzon}},\ and\ \bibinfo {author} {\bibfnamefont {G.}~\bibnamefont {Hetsroni}},\ }\bibfield  {title} {\bibinfo {title} {Detecting coherent patterns in a flume by using piv and ir imaging techniques},\ }\href {https://doi.org/10.1007/s00348-004-0805-3} {\bibfield  {journal} {\bibinfo  {journal} {Exp. Fluids}\ }\textbf {\bibinfo {volume} {37}},\ \bibinfo {pages} {230–236} (\bibinfo {year} {2004}{\natexlab{a}})}\BibitemShut {NoStop}%
\bibitem [{\citenamefont {Antonia}\ \emph {et~al.}(1988)\citenamefont {Antonia}, \citenamefont {Krishnamoorthy},\ and\ \citenamefont {Fulachier}}]{Antonia1988}%
  \BibitemOpen
  \bibfield  {author} {\bibinfo {author} {\bibfnamefont {R.~A.}\ \bibnamefont {Antonia}}, \bibinfo {author} {\bibfnamefont {L.~V.}\ \bibnamefont {Krishnamoorthy}},\ and\ \bibinfo {author} {\bibfnamefont {L.}~\bibnamefont {Fulachier}},\ }\bibfield  {title} {\bibinfo {title} {Correlation between the longitudinal velocity fluctuation and temperature fluctuation in the near-wall region of a turbulent boundary layer},\ }\href@noop {} {\bibfield  {journal} {\bibinfo  {journal} {Int. J. Heat Mass Transf.}\ }\textbf {\bibinfo {volume} {31}},\ \bibinfo {pages} {723} (\bibinfo {year} {1988})}\BibitemShut {NoStop}%
\bibitem [{\citenamefont {Abe}\ \emph {et~al.}(2004)\citenamefont {Abe}, \citenamefont {Kawamura},\ and\ \citenamefont {Matsuo}}]{Abe2004}%
  \BibitemOpen
  \bibfield  {author} {\bibinfo {author} {\bibfnamefont {H.}~\bibnamefont {Abe}}, \bibinfo {author} {\bibfnamefont {H.}~\bibnamefont {Kawamura}},\ and\ \bibinfo {author} {\bibfnamefont {Y.}~\bibnamefont {Matsuo}},\ }\bibfield  {title} {\bibinfo {title} {Surface heat-flux fluctuations in a turbulent channel flow up to {Re}$_\tau$=1020 with {Pr}=0.025 and 0.71},\ }\href {https://doi.org/https://doi.org/10.1016/j.ijheatfluidflow.2004.02.010} {\bibfield  {journal} {\bibinfo  {journal} {International J. Heat Fluid Fl.}\ }\textbf {\bibinfo {volume} {25}},\ \bibinfo {pages} {404} (\bibinfo {year} {2004})}\BibitemShut {NoStop}%
\bibitem [{\citenamefont {Abe}\ and\ \citenamefont {Antonia}(2009)}]{Abe2009}%
  \BibitemOpen
  \bibfield  {author} {\bibinfo {author} {\bibfnamefont {H.}~\bibnamefont {Abe}}\ and\ \bibinfo {author} {\bibfnamefont {R.~A.}\ \bibnamefont {Antonia}},\ }\bibfield  {title} {\bibinfo {title} {Near-wall similarity between velocity and scalar fluctuations in a turbulent channel flow},\ }\href@noop {} {\bibfield  {journal} {\bibinfo  {journal} {Phys. Fluids}\ }\textbf {\bibinfo {volume} {21}},\ \bibinfo {pages} {025109} (\bibinfo {year} {2009})}\BibitemShut {NoStop}%
\bibitem [{\citenamefont {Kim}\ and\ \citenamefont {Lee}(2020)}]{kim2020prediction}%
  \BibitemOpen
  \bibfield  {author} {\bibinfo {author} {\bibfnamefont {J.}~\bibnamefont {Kim}}\ and\ \bibinfo {author} {\bibfnamefont {C.}~\bibnamefont {Lee}},\ }\bibfield  {title} {\bibinfo {title} {Prediction of turbulent heat transfer using convolutional neural networks},\ }\href@noop {} {\bibfield  {journal} {\bibinfo  {journal} {J. Fluid Mech.}\ }\textbf {\bibinfo {volume} {882}},\ \bibinfo {pages} {A18} (\bibinfo {year} {2020})}\BibitemShut {NoStop}%
\bibitem [{\citenamefont {Perry}\ and\ \citenamefont {Hoffmann}(1976)}]{perry1976experimental}%
  \BibitemOpen
  \bibfield  {author} {\bibinfo {author} {\bibfnamefont {A.}~\bibnamefont {Perry}}\ and\ \bibinfo {author} {\bibfnamefont {P.}~\bibnamefont {Hoffmann}},\ }\bibfield  {title} {\bibinfo {title} {An experimental study of turbulent convective heat transfer from a flat plate},\ }\href@noop {} {\bibfield  {journal} {\bibinfo  {journal} {J. Fluid Mech.}\ }\textbf {\bibinfo {volume} {77}},\ \bibinfo {pages} {355} (\bibinfo {year} {1976})}\BibitemShut {NoStop}%
\bibitem [{\citenamefont {Iritani}\ \emph {et~al.}(1983)\citenamefont {Iritani}, \citenamefont {Kasagi},\ and\ \citenamefont {Hirata}}]{iritani1985heat}%
  \BibitemOpen
  \bibfield  {author} {\bibinfo {author} {\bibfnamefont {Y.}~\bibnamefont {Iritani}}, \bibinfo {author} {\bibfnamefont {N.}~\bibnamefont {Kasagi}},\ and\ \bibinfo {author} {\bibfnamefont {M.}~\bibnamefont {Hirata}},\ }\bibfield  {title} {\bibinfo {title} {Heat transfer mechanism and associated turbulence structure in the near-wall region of a turbulent boundary layer},\ }in\ \href@noop {} {\emph {\bibinfo {booktitle} {Turbulent Shear Flows 4: Selected Papers from the Fourth International Symposium on Turbulent Shear Flows, University of Karlsruhe, Karlsruhe, FRG, September 12--14, 1983}}}\ (\bibinfo {organization} {Springer},\ \bibinfo {year} {1983})\ pp.\ \bibinfo {pages} {223--234}\BibitemShut {NoStop}%
\bibitem [{\citenamefont {Kong}\ \emph {et~al.}(2000)\citenamefont {Kong}, \citenamefont {Choi},\ and\ \citenamefont {Lee}}]{kong2000direct}%
  \BibitemOpen
  \bibfield  {author} {\bibinfo {author} {\bibfnamefont {H.}~\bibnamefont {Kong}}, \bibinfo {author} {\bibfnamefont {H.}~\bibnamefont {Choi}},\ and\ \bibinfo {author} {\bibfnamefont {J.~S.}\ \bibnamefont {Lee}},\ }\bibfield  {title} {\bibinfo {title} {Direct numerical simulation of turbulent thermal boundary layers},\ }\href@noop {} {\bibfield  {journal} {\bibinfo  {journal} {Phys. Fluids}\ }\textbf {\bibinfo {volume} {12}},\ \bibinfo {pages} {2555} (\bibinfo {year} {2000})}\BibitemShut {NoStop}%
\bibitem [{\citenamefont {Abe}\ \emph {et~al.}(2001)\citenamefont {Abe}, \citenamefont {Kawamura},\ and\ \citenamefont {Matsuo}}]{Abe2001}%
  \BibitemOpen
  \bibfield  {author} {\bibinfo {author} {\bibfnamefont {H.}~\bibnamefont {Abe}}, \bibinfo {author} {\bibfnamefont {H.}~\bibnamefont {Kawamura}},\ and\ \bibinfo {author} {\bibfnamefont {Y.}~\bibnamefont {Matsuo}},\ }\bibfield  {title} {\bibinfo {title} {Direct numerical simulation of a fully developed turbulent channel flow with respect to the reynolds number dependence},\ }\href@noop {} {\bibfield  {journal} {\bibinfo  {journal} {J. Fluids Eng.}\ }\textbf {\bibinfo {volume} {123}},\ \bibinfo {pages} {382} (\bibinfo {year} {2001})}\BibitemShut {NoStop}%
\bibitem [{\citenamefont {Pirozzoli}\ \emph {et~al.}(2016)\citenamefont {Pirozzoli}, \citenamefont {Bernardini},\ and\ \citenamefont {Orlandi}}]{pirozzoli2016passive}%
  \BibitemOpen
  \bibfield  {author} {\bibinfo {author} {\bibfnamefont {S.}~\bibnamefont {Pirozzoli}}, \bibinfo {author} {\bibfnamefont {M.}~\bibnamefont {Bernardini}},\ and\ \bibinfo {author} {\bibfnamefont {P.}~\bibnamefont {Orlandi}},\ }\bibfield  {title} {\bibinfo {title} {{Passive scalars in turbulent channel flow at high {R}eynolds number}},\ }\href@noop {} {\bibfield  {journal} {\bibinfo  {journal} {J. Fluid Mech.}\ }\textbf {\bibinfo {volume} {788}},\ \bibinfo {pages} {614} (\bibinfo {year} {2016})}\BibitemShut {NoStop}%
\bibitem [{\citenamefont {Alc{\'a}ntara-{\'A}vila}\ and\ \citenamefont {Hoyas}(2021)}]{alcantara2021direct}%
  \BibitemOpen
  \bibfield  {author} {\bibinfo {author} {\bibfnamefont {F.}~\bibnamefont {Alc{\'a}ntara-{\'A}vila}}\ and\ \bibinfo {author} {\bibfnamefont {S.}~\bibnamefont {Hoyas}},\ }\bibfield  {title} {\bibinfo {title} {{Direct numerical simulation of thermal channel flow for medium--high Prandtl numbers up to {Re}$_\tau$= 2000}},\ }\href@noop {} {\bibfield  {journal} {\bibinfo  {journal} {Int. J. Heat Mass Tran.}\ }\textbf {\bibinfo {volume} {176}},\ \bibinfo {pages} {121412} (\bibinfo {year} {2021})}\BibitemShut {NoStop}%
\bibitem [{\citenamefont {Alc{\'a}ntara-{\'A}vila}\ \emph {et~al.}(2021)\citenamefont {Alc{\'a}ntara-{\'A}vila}, \citenamefont {Hoyas},\ and\ \citenamefont {P{\'e}rez-Quiles}}]{alcantara2021directjfm}%
  \BibitemOpen
  \bibfield  {author} {\bibinfo {author} {\bibfnamefont {F.}~\bibnamefont {Alc{\'a}ntara-{\'A}vila}}, \bibinfo {author} {\bibfnamefont {S.}~\bibnamefont {Hoyas}},\ and\ \bibinfo {author} {\bibfnamefont {M.~J.}\ \bibnamefont {P{\'e}rez-Quiles}},\ }\bibfield  {title} {\bibinfo {title} {Direct numerical simulation of thermal channel flow for {Re}$_\tau$= 5000 and {Pr}=0.71},\ }\href@noop {} {\bibfield  {journal} {\bibinfo  {journal} {J. Fluid Mech.}\ }\textbf {\bibinfo {volume} {916}},\ \bibinfo {pages} {A29} (\bibinfo {year} {2021})}\BibitemShut {NoStop}%
\bibitem [{\citenamefont {Miozzi}\ \emph {et~al.}(2024)\citenamefont {Miozzi}, \citenamefont {Schr\"{o}der}, \citenamefont {Schanz}, \citenamefont {Willert}, \citenamefont {Klein},\ and\ \citenamefont {Lemarechal}}]{Miozzi2024}%
  \BibitemOpen
  \bibfield  {author} {\bibinfo {author} {\bibfnamefont {M.}~\bibnamefont {Miozzi}}, \bibinfo {author} {\bibfnamefont {A.}~\bibnamefont {Schr\"{o}der}}, \bibinfo {author} {\bibfnamefont {D.}~\bibnamefont {Schanz}}, \bibinfo {author} {\bibfnamefont {C.~E.}\ \bibnamefont {Willert}}, \bibinfo {author} {\bibfnamefont {C.}~\bibnamefont {Klein}},\ and\ \bibinfo {author} {\bibfnamefont {J.}~\bibnamefont {Lemarechal}},\ }\bibfield  {title} {\bibinfo {title} {Skin-friction from temperature and velocity data around a wall-mounted cube},\ }\bibfield  {journal} {\bibinfo  {journal} {Exp. Fluids}\ }\textbf {\bibinfo {volume} {65}},\ \href {https://doi.org/10.1007/s00348-024-03881-2} {10.1007/s00348-024-03881-2} (\bibinfo {year} {2024})\BibitemShut {NoStop}%
\bibitem [{\citenamefont {Nakamura}\ and\ \citenamefont {Yamada}(2013)}]{nakamura2013}%
  \BibitemOpen
  \bibfield  {author} {\bibinfo {author} {\bibfnamefont {Y.}~\bibnamefont {Nakamura}}\ and\ \bibinfo {author} {\bibfnamefont {H.}~\bibnamefont {Yamada}},\ }\bibfield  {title} {\bibinfo {title} {Turbulence structure identification in a wall-bounded turbulent flow using infrared thermography},\ }\href@noop {} {\bibfield  {journal} {\bibinfo  {journal} {J. Fluid Mech.}\ }\textbf {\bibinfo {volume} {735}},\ \bibinfo {pages} {332} (\bibinfo {year} {2013})}\BibitemShut {NoStop}%
\bibitem [{\citenamefont {Raiola}\ \emph {et~al.}(2017)\citenamefont {Raiola}, \citenamefont {Greco}, \citenamefont {Contino}, \citenamefont {Discetti},\ and\ \citenamefont {Ianiro}}]{RAIOLA2017}%
  \BibitemOpen
  \bibfield  {author} {\bibinfo {author} {\bibfnamefont {M.}~\bibnamefont {Raiola}}, \bibinfo {author} {\bibfnamefont {C.~S.}\ \bibnamefont {Greco}}, \bibinfo {author} {\bibfnamefont {M.}~\bibnamefont {Contino}}, \bibinfo {author} {\bibfnamefont {S.}~\bibnamefont {Discetti}},\ and\ \bibinfo {author} {\bibfnamefont {A.}~\bibnamefont {Ianiro}},\ }\bibfield  {title} {\bibinfo {title} {Towards enabling time-resolved measurements of turbulent convective heat transfer maps with ir thermography and a heated thin foil},\ }\href {https://doi.org/https://doi.org/10.1016/j.ijheatmasstransfer.2016.12.002} {\bibfield  {journal} {\bibinfo  {journal} {Int. J. Heat Mass Tran.}\ }\textbf {\bibinfo {volume} {108}},\ \bibinfo {pages} {199} (\bibinfo {year} {2017})}\BibitemShut {NoStop}%
\bibitem [{\citenamefont {Gurka}\ \emph {et~al.}(2004{\natexlab{b}})\citenamefont {Gurka}, \citenamefont {Liberzon},\ and\ \citenamefont {Hetsroni}}]{gurka2004detecting}%
  \BibitemOpen
  \bibfield  {author} {\bibinfo {author} {\bibfnamefont {R.}~\bibnamefont {Gurka}}, \bibinfo {author} {\bibfnamefont {A.}~\bibnamefont {Liberzon}},\ and\ \bibinfo {author} {\bibfnamefont {G.}~\bibnamefont {Hetsroni}},\ }\bibfield  {title} {\bibinfo {title} {{Detecting coherent patterns in a flume by using PIV and IR imaging techniques}},\ }\href@noop {} {\bibfield  {journal} {\bibinfo  {journal} {Exp. Fluids}\ }\textbf {\bibinfo {volume} {37}},\ \bibinfo {pages} {230} (\bibinfo {year} {2004}{\natexlab{b}})}\BibitemShut {NoStop}%
\bibitem [{\citenamefont {Foroozan}\ \emph {et~al.}(2023)\citenamefont {Foroozan}, \citenamefont {G{\"u}emes}, \citenamefont {Raiola}, \citenamefont {Castellanos}, \citenamefont {Discetti},\ and\ \citenamefont {Ianiro}}]{foroozan2023}%
  \BibitemOpen
  \bibfield  {author} {\bibinfo {author} {\bibfnamefont {F.}~\bibnamefont {Foroozan}}, \bibinfo {author} {\bibfnamefont {A.}~\bibnamefont {G{\"u}emes}}, \bibinfo {author} {\bibfnamefont {M.}~\bibnamefont {Raiola}}, \bibinfo {author} {\bibfnamefont {R.}~\bibnamefont {Castellanos}}, \bibinfo {author} {\bibfnamefont {S.}~\bibnamefont {Discetti}},\ and\ \bibinfo {author} {\bibfnamefont {A.}~\bibnamefont {Ianiro}},\ }\bibfield  {title} {\bibinfo {title} {Synchronized measurement of instantaneous convective heat flux and velocity fields in wall-bounded flows},\ }\href@noop {} {\bibfield  {journal} {\bibinfo  {journal} {Meas. Sci. Tech.}\ }\textbf {\bibinfo {volume} {34}},\ \bibinfo {pages} {125301} (\bibinfo {year} {2023})}\BibitemShut {NoStop}%
\bibitem [{\citenamefont {Cu{\'e}llar}\ \emph {et~al.}(2024{\natexlab{c}})\citenamefont {Cu{\'e}llar}, \citenamefont {Amico}, \citenamefont {Serpieri}, \citenamefont {Cafiero}, \citenamefont {Baars}, \citenamefont {Discetti},\ and\ \citenamefont {Ianiro}}]{cuellar2024measuring}%
  \BibitemOpen
  \bibfield  {author} {\bibinfo {author} {\bibfnamefont {A.}~\bibnamefont {Cu{\'e}llar}}, \bibinfo {author} {\bibfnamefont {E.}~\bibnamefont {Amico}}, \bibinfo {author} {\bibfnamefont {J.}~\bibnamefont {Serpieri}}, \bibinfo {author} {\bibfnamefont {G.}~\bibnamefont {Cafiero}}, \bibinfo {author} {\bibfnamefont {W.~J.}\ \bibnamefont {Baars}}, \bibinfo {author} {\bibfnamefont {S.}~\bibnamefont {Discetti}},\ and\ \bibinfo {author} {\bibfnamefont {A.}~\bibnamefont {Ianiro}},\ }\bibfield  {title} {\bibinfo {title} {Measuring time-resolved heat transfer fluctuations on a heated-thin foil in a turbulent channel airflow},\ }\href@noop {} {\bibfield  {journal} {\bibinfo  {journal} {arXiv preprint arXiv:2410.12778}\ } (\bibinfo {year} {2024}{\natexlab{c}})}\BibitemShut {NoStop}%
\bibitem [{\citenamefont {Baars}\ \emph {et~al.}(2024)\citenamefont {Baars}, \citenamefont {Dacome},\ and\ \citenamefont {Lee}}]{Baars_Dacome_Lee_2024}%
  \BibitemOpen
  \bibfield  {author} {\bibinfo {author} {\bibfnamefont {W.~J.}\ \bibnamefont {Baars}}, \bibinfo {author} {\bibfnamefont {G.}~\bibnamefont {Dacome}},\ and\ \bibinfo {author} {\bibfnamefont {M.}~\bibnamefont {Lee}},\ }\bibfield  {title} {\bibinfo {title} {Reynolds-number scaling of wall-pressure–velocity correlations in wall-bounded turbulence},\ }\href {https://doi.org/10.1017/jfm.2024.46} {\bibfield  {journal} {\bibinfo  {journal} {J. Fluid Mech.}\ }\textbf {\bibinfo {volume} {981}},\ \bibinfo {pages} {A15} (\bibinfo {year} {2024})}\BibitemShut {NoStop}%
\bibitem [{\citenamefont {Chauhan}\ \emph {et~al.}(2009)\citenamefont {Chauhan}, \citenamefont {Monkewitz},\ and\ \citenamefont {Nagib}}]{chauhan:2009a}%
  \BibitemOpen
  \bibfield  {author} {\bibinfo {author} {\bibfnamefont {K.~A.}\ \bibnamefont {Chauhan}}, \bibinfo {author} {\bibfnamefont {P.~A.}\ \bibnamefont {Monkewitz}},\ and\ \bibinfo {author} {\bibfnamefont {H.~M.}\ \bibnamefont {Nagib}},\ }\bibfield  {title} {\bibinfo {title} {Criteria for assessing experiments in zero pressure gradient boundary layers},\ }\href@noop {} {\bibfield  {journal} {\bibinfo  {journal} {Fluid Dyn. Res.}\ }\textbf {\bibinfo {volume} {41}} (\bibinfo {year} {2009})},\ \bibinfo {note} {021404}\BibitemShut {NoStop}%
\bibitem [{\citenamefont {Sillero}\ \emph {et~al.}(2014)\citenamefont {Sillero}, \citenamefont {Jiménez},\ and\ \citenamefont {Moser}}]{Sillero2014}%
  \BibitemOpen
  \bibfield  {author} {\bibinfo {author} {\bibfnamefont {J.~A.}\ \bibnamefont {Sillero}}, \bibinfo {author} {\bibfnamefont {J.}~\bibnamefont {Jiménez}},\ and\ \bibinfo {author} {\bibfnamefont {R.~D.}\ \bibnamefont {Moser}},\ }\bibfield  {title} {\bibinfo {title} {{Two-point statistics for turbulent boundary layers and channels at Reynolds numbers up to $\delta^+ \approx 2000$}},\ }\bibfield  {journal} {\bibinfo  {journal} {Phys. Fluids}\ }\textbf {\bibinfo {volume} {26}},\ \href {https://doi.org/10.1063/1.4899259} {10.1063/1.4899259} (\bibinfo {year} {2014})\BibitemShut {NoStop}%
\bibitem [{\citenamefont {Nakamura}(2009)}]{NAKAMURA2009}%
  \BibitemOpen
  \bibfield  {author} {\bibinfo {author} {\bibfnamefont {H.}~\bibnamefont {Nakamura}},\ }\bibfield  {title} {\bibinfo {title} {Frequency response and spatial resolution of a thin foil for heat transfer measurements using infrared thermography},\ }\href {https://doi.org/https://doi.org/10.1016/j.ijheatmasstransfer.2009.04.019} {\bibfield  {journal} {\bibinfo  {journal} {Int. J. Heat Mass Tran.}\ }\textbf {\bibinfo {volume} {52}},\ \bibinfo {pages} {5040} (\bibinfo {year} {2009})}\BibitemShut {NoStop}%
\bibitem [{\citenamefont {Cattell}(1966)}]{Cattell1966}%
  \BibitemOpen
  \bibfield  {author} {\bibinfo {author} {\bibfnamefont {R.~B.}\ \bibnamefont {Cattell}},\ }\bibfield  {title} {\bibinfo {title} {The scree test for the number of factors},\ }\href {https://doi.org/10.1207/s15327906mbr0102\_10} {\bibfield  {journal} {\bibinfo  {journal} {Multiv. Behav. Res.}\ }\textbf {\bibinfo {volume} {1}},\ \bibinfo {pages} {245} (\bibinfo {year} {1966})}\BibitemShut {NoStop}%
\bibitem [{\citenamefont {Moffat}(1988)}]{MOFFAT19883}%
  \BibitemOpen
  \bibfield  {author} {\bibinfo {author} {\bibfnamefont {R.~J.}\ \bibnamefont {Moffat}},\ }\bibfield  {title} {\bibinfo {title} {Describing the uncertainties in experimental results},\ }\href@noop {} {\bibfield  {journal} {\bibinfo  {journal} {Exp. Therm. Fluid Sci.}\ }\textbf {\bibinfo {volume} {1}},\ \bibinfo {pages} {3} (\bibinfo {year} {1988})}\BibitemShut {NoStop}%
\bibitem [{\citenamefont {Meola}\ \emph {et~al.}(2015)\citenamefont {Meola}, \citenamefont {Boccardi},\ and\ \citenamefont {Carlomagno}}]{meola2015measurements}%
  \BibitemOpen
  \bibfield  {author} {\bibinfo {author} {\bibfnamefont {C.}~\bibnamefont {Meola}}, \bibinfo {author} {\bibfnamefont {S.}~\bibnamefont {Boccardi}},\ and\ \bibinfo {author} {\bibfnamefont {G.~M.}\ \bibnamefont {Carlomagno}},\ }\bibfield  {title} {\bibinfo {title} {Measurements of very small temperature variations with lwir qwip infrared camera},\ }\href@noop {} {\bibfield  {journal} {\bibinfo  {journal} {Infrared Phys. Techn.}\ }\textbf {\bibinfo {volume} {72}},\ \bibinfo {pages} {195} (\bibinfo {year} {2015})}\BibitemShut {NoStop}%
\bibitem [{\citenamefont {Howard}\ and\ \citenamefont {Abel}(1982)}]{howard1982narcissus}%
  \BibitemOpen
  \bibfield  {author} {\bibinfo {author} {\bibfnamefont {J.~W.}\ \bibnamefont {Howard}}\ and\ \bibinfo {author} {\bibfnamefont {I.~R.}\ \bibnamefont {Abel}},\ }\bibfield  {title} {\bibinfo {title} {Narcissus: reflections on retroreflections in thermal imaging systems},\ }\href@noop {} {\bibfield  {journal} {\bibinfo  {journal} {Appl. Optics}\ }\textbf {\bibinfo {volume} {21}},\ \bibinfo {pages} {3393} (\bibinfo {year} {1982})}\BibitemShut {NoStop}%
\bibitem [{\citenamefont {Liu}\ and\ \citenamefont {Gayme}(2020)}]{Liu_Gayme_2020}%
  \BibitemOpen
  \bibfield  {author} {\bibinfo {author} {\bibfnamefont {C.}~\bibnamefont {Liu}}\ and\ \bibinfo {author} {\bibfnamefont {D.~F.}\ \bibnamefont {Gayme}},\ }\bibfield  {title} {\bibinfo {title} {An input–output based analysis of convective velocity in turbulent channels},\ }\href {https://doi.org/10.1017/jfm.2020.48} {\bibfield  {journal} {\bibinfo  {journal} {J. Fluid Mech.}\ }\textbf {\bibinfo {volume} {888}},\ \bibinfo {pages} {A32} (\bibinfo {year} {2020})}\BibitemShut {NoStop}%
\bibitem [{\citenamefont {McAdams}\ \emph {et~al.}(1926)\citenamefont {McAdams}, \citenamefont {Sherwood},\ and\ \citenamefont {Turner}}]{mcadams1926heat}%
  \BibitemOpen
  \bibfield  {author} {\bibinfo {author} {\bibfnamefont {W.}~\bibnamefont {McAdams}}, \bibinfo {author} {\bibfnamefont {T.}~\bibnamefont {Sherwood}},\ and\ \bibinfo {author} {\bibfnamefont {R.}~\bibnamefont {Turner}},\ }\bibfield  {title} {\bibinfo {title} {Heat transmission from condensing steam to water in surface condensers and feedwater heaters},\ }\href@noop {} {\bibfield  {journal} {\bibinfo  {journal} {J. Fluids Eng.}\ }\textbf {\bibinfo {volume} {48}},\ \bibinfo {pages} {1233} (\bibinfo {year} {1926})}\BibitemShut {NoStop}%
\bibitem [{\citenamefont {Kline}\ \emph {et~al.}(1967)\citenamefont {Kline}, \citenamefont {Reynolds}, \citenamefont {Schraub},\ and\ \citenamefont {Runstadler}}]{Kline1967}%
  \BibitemOpen
  \bibfield  {author} {\bibinfo {author} {\bibfnamefont {S.~J.}\ \bibnamefont {Kline}}, \bibinfo {author} {\bibfnamefont {W.~C.}\ \bibnamefont {Reynolds}}, \bibinfo {author} {\bibfnamefont {F.~A.}\ \bibnamefont {Schraub}},\ and\ \bibinfo {author} {\bibfnamefont {P.~W.}\ \bibnamefont {Runstadler}},\ }\bibfield  {title} {\bibinfo {title} {The structure of turbulent boundary layers},\ }\href {https://doi.org/10.1017/S0022112067001740} {\bibfield  {journal} {\bibinfo  {journal} {J. Fluid Mech.}\ }\textbf {\bibinfo {volume} {30}},\ \bibinfo {pages} {741–773} (\bibinfo {year} {1967})}\BibitemShut {NoStop}%
\bibitem [{\citenamefont {Smith}\ and\ \citenamefont {Metzler}(1983)}]{Smith_Metzler_1983}%
  \BibitemOpen
  \bibfield  {author} {\bibinfo {author} {\bibfnamefont {C.~R.}\ \bibnamefont {Smith}}\ and\ \bibinfo {author} {\bibfnamefont {S.~P.}\ \bibnamefont {Metzler}},\ }\bibfield  {title} {\bibinfo {title} {The characteristics of low-speed streaks in the near-wall region of a turbulent boundary layer},\ }\href {https://doi.org/10.1017/S0022112083000634} {\bibfield  {journal} {\bibinfo  {journal} {J. Fluid Mech.}\ }\textbf {\bibinfo {volume} {129}},\ \bibinfo {pages} {27–54} (\bibinfo {year} {1983})}\BibitemShut {NoStop}%
\bibitem [{\citenamefont {Klewicki}\ \emph {et~al.}(1995)\citenamefont {Klewicki}, \citenamefont {Metzger}, \citenamefont {Kelner},\ and\ \citenamefont {Thurlow}}]{Klewicki1995}%
  \BibitemOpen
  \bibfield  {author} {\bibinfo {author} {\bibfnamefont {J.~C.}\ \bibnamefont {Klewicki}}, \bibinfo {author} {\bibfnamefont {M.~M.}\ \bibnamefont {Metzger}}, \bibinfo {author} {\bibfnamefont {E.}~\bibnamefont {Kelner}},\ and\ \bibinfo {author} {\bibfnamefont {E.~M.}\ \bibnamefont {Thurlow}},\ }\bibfield  {title} {\bibinfo {title} {{Viscous sublayer flow visualizations at {$\mathbf{R}_\theta\approx1.500.000$}}},\ }\href {https://doi.org/10.1063/1.868763} {\bibfield  {journal} {\bibinfo  {journal} {Phys. Fluids}\ }\textbf {\bibinfo {volume} {7}},\ \bibinfo {pages} {857} (\bibinfo {year} {1995})}\BibitemShut {NoStop}%
\bibitem [{\citenamefont {Smits}\ \emph {et~al.}(2011)\citenamefont {Smits}, \citenamefont {McKeon},\ and\ \citenamefont {Marusic}}]{Smits2011}%
  \BibitemOpen
  \bibfield  {author} {\bibinfo {author} {\bibfnamefont {A.~J.}\ \bibnamefont {Smits}}, \bibinfo {author} {\bibfnamefont {B.~J.}\ \bibnamefont {McKeon}},\ and\ \bibinfo {author} {\bibfnamefont {I.}~\bibnamefont {Marusic}},\ }\bibfield  {title} {\bibinfo {title} {High–{R}eynolds number wall turbulence},\ }\href {https://doi.org/https://doi.org/10.1146/annurev-fluid-122109-160753} {\bibfield  {journal} {\bibinfo  {journal} {Annu. Rev. Fluid Mech.}\ }\textbf {\bibinfo {volume} {43}},\ \bibinfo {pages} {353} (\bibinfo {year} {2011})}\BibitemShut {NoStop}%
\bibitem [{\citenamefont {Li}\ \emph {et~al.}(2009)\citenamefont {Li}, \citenamefont {Schlatter}, \citenamefont {Brandt},\ and\ \citenamefont {Henningson}}]{li2009dns}%
  \BibitemOpen
  \bibfield  {author} {\bibinfo {author} {\bibfnamefont {Q.}~\bibnamefont {Li}}, \bibinfo {author} {\bibfnamefont {P.}~\bibnamefont {Schlatter}}, \bibinfo {author} {\bibfnamefont {L.}~\bibnamefont {Brandt}},\ and\ \bibinfo {author} {\bibfnamefont {D.~S.}\ \bibnamefont {Henningson}},\ }\bibfield  {title} {\bibinfo {title} {{DNS of a spatially developing turbulent boundary layer with passive scalar transport}},\ }\href@noop {} {\bibfield  {journal} {\bibinfo  {journal} {International J. Heat Fluid Fl.}\ }\textbf {\bibinfo {volume} {30}},\ \bibinfo {pages} {916} (\bibinfo {year} {2009})}\BibitemShut {NoStop}%
\bibitem [{\citenamefont {Balasubramanian}\ \emph {et~al.}(2023)\citenamefont {Balasubramanian}, \citenamefont {Guastoni}, \citenamefont {Schlatter},\ and\ \citenamefont {Vinuesa}}]{balasubramanian2023direct}%
  \BibitemOpen
  \bibfield  {author} {\bibinfo {author} {\bibfnamefont {A.~G.}\ \bibnamefont {Balasubramanian}}, \bibinfo {author} {\bibfnamefont {L.}~\bibnamefont {Guastoni}}, \bibinfo {author} {\bibfnamefont {P.}~\bibnamefont {Schlatter}},\ and\ \bibinfo {author} {\bibfnamefont {R.}~\bibnamefont {Vinuesa}},\ }\bibfield  {title} {\bibinfo {title} {{Direct numerical simulation of a zero-pressure-gradient turbulent boundary layer with passive scalars up to Prandtl number Pr= 6}},\ }\href@noop {} {\bibfield  {journal} {\bibinfo  {journal} {J. Fluid Mech.}\ }\textbf {\bibinfo {volume} {974}},\ \bibinfo {pages} {A49} (\bibinfo {year} {2023})}\BibitemShut {NoStop}%
\bibitem [{\citenamefont {Wills}(1964)}]{wills_1964}%
  \BibitemOpen
  \bibfield  {author} {\bibinfo {author} {\bibfnamefont {J.~A.~B.}\ \bibnamefont {Wills}},\ }\bibfield  {title} {\bibinfo {title} {On convection velocities in turbulent shear flows},\ }\href {https://doi.org/10.1017/S002211206400132X} {\bibfield  {journal} {\bibinfo  {journal} {J. Fluid Mech.}\ }\textbf {\bibinfo {volume} {20}},\ \bibinfo {pages} {417–432} (\bibinfo {year} {1964})}\BibitemShut {NoStop}%
\bibitem [{\citenamefont {Choi}\ and\ \citenamefont {Moin}(1990)}]{Choi1990}%
  \BibitemOpen
  \bibfield  {author} {\bibinfo {author} {\bibfnamefont {H.}~\bibnamefont {Choi}}\ and\ \bibinfo {author} {\bibfnamefont {P.}~\bibnamefont {Moin}},\ }\bibfield  {title} {\bibinfo {title} {On the space-time characteristics of wall-pressure fluctuations},\ }\href {https://doi.org/10.1063/1.857593} {\bibfield  {journal} {\bibinfo  {journal} {Phys. Fluids A: Fluid}\ }\textbf {\bibinfo {volume} {2}},\ \bibinfo {pages} {1450–1460} (\bibinfo {year} {1990})}\BibitemShut {NoStop}%
\bibitem [{\citenamefont {Kim}\ and\ \citenamefont {Hussain}(1993)}]{Kim1993}%
  \BibitemOpen
  \bibfield  {author} {\bibinfo {author} {\bibfnamefont {J.}~\bibnamefont {Kim}}\ and\ \bibinfo {author} {\bibfnamefont {F.}~\bibnamefont {Hussain}},\ }\bibfield  {title} {\bibinfo {title} {Propagation velocity of perturbations in turbulent channel flow},\ }\href {https://doi.org/10.1063/1.858653} {\bibfield  {journal} {\bibinfo  {journal} {Phys. Fluids A: Fluid}\ }\textbf {\bibinfo {volume} {5}},\ \bibinfo {pages} {695–706} (\bibinfo {year} {1993})}\BibitemShut {NoStop}%
\bibitem [{\citenamefont {Jeon}\ \emph {et~al.}(1999)\citenamefont {Jeon}, \citenamefont {Choi}, \citenamefont {Yoo},\ and\ \citenamefont {Moin}}]{Jeon1999}%
  \BibitemOpen
  \bibfield  {author} {\bibinfo {author} {\bibfnamefont {S.}~\bibnamefont {Jeon}}, \bibinfo {author} {\bibfnamefont {H.}~\bibnamefont {Choi}}, \bibinfo {author} {\bibfnamefont {J.~Y.}\ \bibnamefont {Yoo}},\ and\ \bibinfo {author} {\bibfnamefont {P.}~\bibnamefont {Moin}},\ }\bibfield  {title} {\bibinfo {title} {Space–time characteristics of the wall shear-stress fluctuations in a low-reynolds-number channel flow},\ }\href {https://doi.org/10.1063/1.870166} {\bibfield  {journal} {\bibinfo  {journal} {Phys. Fluids}\ }\textbf {\bibinfo {volume} {11}},\ \bibinfo {pages} {3084–3094} (\bibinfo {year} {1999})}\BibitemShut {NoStop}%
\bibitem [{\citenamefont {Del~{\'A}lamo}\ and\ \citenamefont {Jim{\'e}nez}(2009)}]{DELÁLAMO_JIMÉNEZ_2009}%
  \BibitemOpen
  \bibfield  {author} {\bibinfo {author} {\bibfnamefont {J.}~\bibnamefont {Del~{\'A}lamo}}\ and\ \bibinfo {author} {\bibfnamefont {J.}~\bibnamefont {Jim{\'e}nez}},\ }\bibfield  {title} {\bibinfo {title} {Estimation of turbulent convection velocities and corrections to {T}aylor’s approximation},\ }\href {https://doi.org/10.1017/S0022112009991029} {\bibfield  {journal} {\bibinfo  {journal} {J. Fluid Mech.}\ }\textbf {\bibinfo {volume} {640}},\ \bibinfo {pages} {5–26} (\bibinfo {year} {2009})}\BibitemShut {NoStop}%
\bibitem [{\citenamefont {Moin}(2009)}]{MOIN_2009}%
  \BibitemOpen
  \bibfield  {author} {\bibinfo {author} {\bibfnamefont {P.}~\bibnamefont {Moin}},\ }\bibfield  {title} {\bibinfo {title} {Revisiting {T}aylor’s hypothesis},\ }\href {https://doi.org/10.1017/S0022112009992126} {\bibfield  {journal} {\bibinfo  {journal} {J. Fluid Mech.}\ }\textbf {\bibinfo {volume} {640}},\ \bibinfo {pages} {1–4} (\bibinfo {year} {2009})}\BibitemShut {NoStop}%
\bibitem [{\citenamefont {Goldschmidt}\ \emph {et~al.}(1981)\citenamefont {Goldschmidt}, \citenamefont {Young},\ and\ \citenamefont {Ott}}]{goldschmidt_young_ott_1981}%
  \BibitemOpen
  \bibfield  {author} {\bibinfo {author} {\bibfnamefont {V.~W.}\ \bibnamefont {Goldschmidt}}, \bibinfo {author} {\bibfnamefont {M.~F.}\ \bibnamefont {Young}},\ and\ \bibinfo {author} {\bibfnamefont {E.~S.}\ \bibnamefont {Ott}},\ }\bibfield  {title} {\bibinfo {title} {Turbulent convective velocities (broadband and wavenumber dependent) in a plane jet},\ }\href {https://doi.org/10.1017/S0022112081003236} {\bibfield  {journal} {\bibinfo  {journal} {J. Fluid Mech.}\ }\textbf {\bibinfo {volume} {105}},\ \bibinfo {pages} {327–345} (\bibinfo {year} {1981})}\BibitemShut {NoStop}%
\bibitem [{\citenamefont {Krogstad}\ \emph {et~al.}(1998)\citenamefont {Krogstad}, \citenamefont {Kaspersen},\ and\ \citenamefont {Rimestad}}]{Krogstad1998ConvectionVI}%
  \BibitemOpen
  \bibfield  {author} {\bibinfo {author} {\bibfnamefont {P.}~\bibnamefont {Krogstad}}, \bibinfo {author} {\bibfnamefont {J.~H.}\ \bibnamefont {Kaspersen}},\ and\ \bibinfo {author} {\bibfnamefont {S.}~\bibnamefont {Rimestad}},\ }\bibfield  {title} {\bibinfo {title} {Convection velocities in a turbulent boundary layer},\ }\href@noop {} {\bibfield  {journal} {\bibinfo  {journal} {Phys. Fluids}\ }\textbf {\bibinfo {volume} {10}},\ \bibinfo {pages} {949} (\bibinfo {year} {1998})}\BibitemShut {NoStop}%
\bibitem [{\citenamefont {del {\'A}lamo}\ and\ \citenamefont {Jim{\'e}nez}(2004)}]{delalamo2004}%
  \BibitemOpen
  \bibfield  {author} {\bibinfo {author} {\bibfnamefont {J.}~\bibnamefont {del {\'A}lamo}}\ and\ \bibinfo {author} {\bibfnamefont {J.}~\bibnamefont {Jim{\'e}nez}},\ }\bibfield  {title} {\bibinfo {title} {Scaling of the energy spectra of turbulent channels},\ }\href@noop {} {\bibfield  {journal} {\bibinfo  {journal} {J. Fluid Mech.}\ }\textbf {\bibinfo {volume} {500}},\ \bibinfo {pages} {135} (\bibinfo {year} {2004})}\BibitemShut {NoStop}%
\end{thebibliography}%
